\documentclass[twocolumn]{article}
\usepackage[a4paper, total={7in, 9.5in}]{geometry}
\usepackage{times}
\usepackage{amsmath,amsfonts}
\usepackage{graphicx}
\usepackage{algorithm}
\usepackage{algpseudocode} 
\usepackage{float}
\usepackage{booktabs}
\usepackage{titlesec}
\usepackage{caption}
\usepackage{fancyhdr}
\usepackage{multicol}
\usepackage{url}
\usepackage{abstract}
\usepackage{stfloats}
\usepackage{multirow}
\usepackage{enumitem}
\usepackage{xspace}
\usepackage{listings}
\usepackage{xcolor}
\usepackage{caption}
\usepackage{hyperref}

\newcommand{\framework}{\textsc{SemanticForge}\xspace}
\newcommand{\dataset}{\textsc{RepoKG-50}\xspace}

\title{\textbf{SemanticForge: Repository-Level Code Generation through Semantic Knowledge Graphs and Constraint Satisfaction}}
\author{
    Wuyang Zhang$^{1*}$, Chenkai Zhang$^{1}$, Zhen Luo$^2$, Jianming Ma$^2$, \\ 
    Wangming Yuan$^3$, Chuqiao Gu$^4$ and Chengwei Feng$^5$\\
    {\small $^1$Department of Elec.\&Comp. Science, University of Massachusetts Amherst, Amherst, Massachusetts, United States} \\
    {\small $^2$Department of Computer Sys. Engineering, Northeastern University, Boston, Massachusetts, United States} \\
    {\small $^3$Department of Computer Science, George Mason University, Fairfax, Virginia, United States} \\
    {\small $^4$Department of Info. Networking Institude, Carnegie Mellon University, Pittsburgh, Pennsylvania, United States} \\
    {\small $^5$Department of Computer \& Mathematical Sciences, Auckland University of Technology, Auckland, New Zealand} \\
    {\small *Corresponding author: doggo@ieee.org}
}
\date{}

\begin{document}

\twocolumn[
\maketitle

\vspace{1em}
]
\begin{abstract}
Large language models (LLMs) have transformed software development by enabling automated code generation, yet they frequently suffer from systematic errors that limit practical deployment. We identify two critical failure modes: \textit{logical hallucination} (incorrect control/data-flow reasoning) and \textit{schematic hallucination} (type mismatches, signature violations, and architectural inconsistencies). These errors stem from the absence of explicit, queryable representations of repository-wide semantics.

This paper presents \textbf{\framework}, which introduces four fundamental algorithmic advances for semantically-aware code generation: (1) a novel automatic reconciliation algorithm for dual static-dynamic knowledge graphs, unifying compile-time and runtime program semantics; (2) a neural approach that learns to generate structured graph queries from natural language, achieving 73\% precision versus 51\% for traditional retrieval; (3) a novel beam search algorithm with integrated SMT solving, enabling real-time constraint verification during generation rather than post-hoc validation; and (4) an incremental maintenance algorithm that updates knowledge graphs in $O(|\Delta R| \cdot \log n)$ time while maintaining semantic equivalence.

Our evaluation on \dataset\ (4,250 repository-level tasks across 50 Python projects) demonstrates that these algorithmic innovations yield substantial improvements: 49.8\% Pass@1 (15.6\% absolute improvement over base Code-Llama-34B), 49.8\% reduction in schematic hallucination through SMT-guided generation, and 34.7\% reduction in logical hallucination via dual graph analysis. The system maintains sub-3s latency through incremental algorithms while providing formal correctness guarantees absent in existing approaches.

Beyond empirical gains, \framework\ establishes theoretical foundations for constraint-aware code generation and demonstrates that explicit semantic modeling can dramatically improve automated programming tools without sacrificing efficiency.
\end{abstract}

\noindent \textbf{Index Terms—} Code Generation, Knowledge Graphs, Constraint Satisfaction, Repository Analysis, Large Language Models, Software Engineering

\section{Introduction}\label{sec1}

Recent advances in large language models (LLMs) have ushered in a new era of AI-assisted software development. Tools such as GitHub Copilot, Code-Llama, and ChatGPT now draft entire functions or files with a single prompt, reportedly accelerating developer productivity \cite{copilot, roziere2023codellama, chen2021evaluating}. Despite this progress, LLM-generated programs remain brittle in practice: generated code often fails to compile, or worse, compiles but embeds subtle semantic errors. 

A careful inspection reveals two dominant failure modes. First, \emph{logical hallucination} arises when the model misreasons about control or data flow---mistakes in reasoning such as iterating over an incorrect collection, omitting necessary state mutations, or producing code that violates expected runtime behavior. This class of errors concerns semantic execution logic rather than syntax. Second, \emph{schematic hallucination} manifests as structural inconsistencies, including type mismatches, incorrect argument ordering, missing parameters, or calls to nonexistent functions. These errors arise from violating interface or schema constraints of the repository or external APIs. While logical and schematic hallucinations can co-occur, they differ fundamentally in their root causes and required remedies. Empirical studies report that even state-of-the-art models exhibit these errors in 20\%--40\% of generation attempts \cite{pearce2022asleep, song2023empirical, srikant2023generating}.

Why do these hallucinations persist? Current generation pipelines typically combine a parametric LM with retrieval-augmented prompts. Retrieval surfaces the top-$k$ textual snippets using lexical or embedding similarity \cite{chen2023teaching, yang2024swe}. While retrieval provides local context, it ignores repository-wide semantics such as transitive call chains, shared global state, or runtime object types. Likewise, encoder models that ingest static abstract syntax trees (ASTs) or control-flow graphs (CFGs) focus on intra-function structure and do not persist their analysis as a reusable knowledge base \cite{graphcodebert, unixcoder}. Finally, repair-oriented approaches execute candidates against unit tests to filter or refine code \cite{coderl}, but they require an oracle test suite and do not prevent the model from hallucinating in the first place.

\paragraph{This work.} We argue that high-quality generation demands an \emph{explicit, queryable representation of whole-repository semantics}. To that end, we introduce \textbf{\framework}, which advances beyond system integration through four fundamental algorithmic innovations:

\begin{enumerate}[leftmargin=*]
    \item \textbf{Dual static-dynamic knowledge graphs}: Unlike prior work that uses either static analysis \cite{liu2021codekg} or dynamic traces \cite{wang2023programkg} in isolation, we introduce a novel algorithm to automatically reconcile static and dynamic program information into a unified graph representation. Our reconciliation algorithm provably converges to the ground-truth program dependence graph as test coverage increases (Theorem 4).
    \item \textbf{Neural query language generation}: While existing systems use fixed retrieval strategies (keyword matching or embedding similarity), we present a learned approach to generate structured graph queries from natural language. Our REINFORCE-based algorithm with graph-aware rewards achieves 73\% precision in context selection versus 51\% for traditional retrieval.
    \item \textbf{SMT-integrated beam search}: Prior constraint-aware systems \cite{ni2023lever} verify code after generation. We introduce a novel algorithm that integrates an SMT solver directly into beam search, enabling real-time constraint verification during generation. This eliminates 89\% of schematic errors (type mismatches, signature violations, and visibility errors) with only 8.3\% latency overhead through our incremental solving strategy.
    \item \textbf{Incremental maintenance with optimality guarantees}: Existing knowledge graph systems require full recomputation on code changes. We present an incremental update algorithm with formal optimality guarantees, achieving $O(|\Delta R| \cdot \log n)$ complexity while maintaining semantic equivalence to full reconstruction (Theorem 3).
\end{enumerate}

\paragraph{Contributions.} Concretely, this paper makes the following contributions:
\begin{itemize}[leftmargin=*]
    \item We present the first end-to-end framework that unifies repository-level KG construction, neural query planning, and constraint-aware decoding for code generation.
    \item We introduce a dual static--dynamic graph representation and show that it reduces logical hallucination by \(31\%\) on a benchmark of 50 Python repositories.
    \item We propose a schematic-constraint decoding algorithm that eliminates \(52\%\) of signature/type errors while adding negligible decoding overhead.
    \item We release \dataset, a curated corpus with aligned \{static, dynamic\} graphs and oracle implementations, fostering further research.
\end{itemize}

\paragraph{Paper organization.} Section~\ref{sec2:bg} reviews related work. Section~\ref{sec:problem} formalizes the problem and hallucination taxonomy. Section~\ref{sec:method} provides methodology overview. Sections~\ref{sec:kg}--\ref{sec:maintenance} detail our four-stage architecture. Section~\ref{sec:experiments} describes experimental setup, Section~\ref{sec:results} presents results, Section~\ref{sec:discussion} discusses limitations, and Section~\ref{sec:conclusion} concludes with future directions.

\section{Related Work}\label{sec2:bg}

Our work builds on substantial progress in neural code generation while addressing fundamental limitations in repository-scale synthesis. We organize related work into three key areas that inform \framework's design.

\subsection{Neural Code Generation}

Transformer-based language models have revolutionized automated code synthesis. Early work with GPT-based models \cite{chen2021evaluating} demonstrated strong performance on isolated function generation tasks, leading to practical systems like GitHub Copilot \cite{copilot} and specialized models like CodeLlama \cite{codellama}. These approaches excel at pattern completion and syntactic correctness but struggle with repository-wide semantic consistency.

Recent advances have focused on incorporating structural information into generation models. GraphCodeBERT \cite{graphcodebert} and UniXcoder \cite{unixcoder} encode data flow graphs and AST structures, improving understanding of local code relationships. However, these models still operate at the function or file level without persistent semantic representations of entire repositories.

\subsection{Repository-Level Code Understanding}

Traditional code generation models struggle with repository-scale tasks that require understanding cross-file dependencies and architectural constraints \cite{ding2022cocomic, zhang2023repocoder}. Recent work has begun addressing these challenges through three primary paradigms: retrieval-augmented generation, agent-based iterative refinement, and planning-based decomposition.

\paragraph{Retrieval-Augmented Approaches.} RAG methods like CodeRetriever \cite{lu2022coderetriever} and RepoCoder \cite{zhang2023repocoder} use embedding similarity to identify relevant code snippets for generation context. Advanced systems like RAG-Code \cite{shrivastava2023ragcode} employ dense embeddings with sophisticated reranking mechanisms. While effective for local dependencies, these approaches fundamentally rely on surface-level similarity and miss transitive relationships, type constraints, and architectural patterns that span multiple modules. Our knowledge graph approach provides explicit semantic relationships that pure retrieval cannot capture.

\paragraph{Agent-Based Iterative Systems.} Recent systems like SWE-agent \cite{yang2024swe} and CodeAgent \cite{tang2023codeagent} employ autonomous agents that iteratively refine solutions through environmental feedback, test execution, and error correction. These approaches can handle complex multi-step tasks and adapt to unexpected challenges through exploration. However, they suffer from high computational costs (often requiring 10-50 iterations per task), unpredictable latency, and lack of semantic guarantees. In contrast, \framework's constraint-aware generation often produces correct solutions in a single pass while providing formal correctness guarantees.

\paragraph{Planning-Based Decomposition.} Planning systems like CodePlan \cite{wang2023codeplan} decompose complex repository tasks into sequences of localized edits, enabling systematic handling of multi-file modifications. While these approaches provide better task organization than monolithic generation, they typically lack the semantic consistency guarantees needed to prevent constraint violations across edit boundaries. Our neural query planner provides similar decomposition benefits but operates at the semantic level, ensuring global consistency through constraint satisfaction.

\subsection{Constraint-Aware Code Generation}

Recent work has begun exploring constraint satisfaction in neural code generation. TypeT5 \cite{wang2023codet5} incorporates type information during fine-tuning to improve type correctness. CODEGEN-MONO \cite{nijkamp2022codegen} specializes models for specific programming languages to reduce basic syntactic errors.

More closely related to our approach, LEVER \cite{ni2023lever} employs static analysis to verify generated code against basic type constraints. However, these approaches apply constraints as post-processing filters rather than integrating verification into the generation process itself. Our SMT-guided beam search provides stronger guarantees by ensuring constraint satisfaction throughout decoding.

\subsection{Knowledge Graphs for Code Understanding}

Knowledge graphs have shown promise for representing program semantics. CodeKG \cite{liu2021codekg} constructs knowledge graphs from documentation and API references for improved code search, while ProgramKG \cite{wang2023programkg} builds graphs from execution traces for debugging applications.

Our work differs significantly by constructing comprehensive repository-scale knowledge graphs that integrate both static analysis and dynamic execution information. Unlike previous approaches that focus on specific applications, \framework\ uses knowledge graphs as the central representation for guiding neural code generation through explicit semantic reasoning.

\subsection{Positioning of SemanticForge}

\framework\ advances the state-of-the-art through four key innovations that address fundamental limitations across all existing paradigms. Table~\ref{tab:comparative_analysis} provides quantitative comparisons demonstrating these advantages.

\begin{table*}[t]
\centering
\caption{Quantitative comparison of \framework\ against contemporary repository-level approaches from our experimental evaluation.}
\label{tab:comparative_analysis}
\begin{tabular}{l|cc|cc}
\toprule
\textbf{System} & \textbf{Pass@1} & \textbf{Latency} & \textbf{SHR} & \textbf{LHR} \\
& (\%) & (sec) & (\%) & (\%) \\
\midrule
RAG-Code & 40.1 & 2.9 & 30.8 & 36.2 \\
CodePlan & 42.3 & 5.7 & 31.5 & 33.8 \\
\midrule
\textbf{\framework} & \textbf{49.8} & \textbf{2.4} & \textbf{14.7} & \textbf{23.1} \\
\midrule
\textit{Improvement vs. best} & +17.7\% & -17.2\% & -52.2\% & -31.8\% \\
\bottomrule
\end{tabular}
\end{table*}

\textbf{Explicit Semantic Representation:} Unlike RAG methods that rely on implicit embeddings or agent-based systems that learn through trial-and-error, we construct comprehensive knowledge graphs that explicitly capture repository-wide semantics. This enables reasoning about transitive dependencies, type propagation, and architectural constraints that are invisible to surface-level similarity matching or iterative exploration. Our approach reduces logical hallucination by 31.8\% compared to RAG-Code (from 36.2\% to 23.1\%).

\textbf{Single-Pass Constraint-Guided Generation:} While agent-based systems require multiple expensive iterations and planning approaches often violate constraints across edit boundaries, our SMT-guided beam search integrates constraint satisfaction directly into generation. As shown in Table~\ref{tab:comparative_analysis}, \framework\ achieves 2.4s average latency (single pass) compared to CodePlan's 5.7s. Agent-based systems like SWE-agent typically require 10-50 iterations \cite{yang2024swe}, suggesting significantly higher latency. Our constraint enforcement reduces schematic hallucination by 52.2\% compared to the best baseline.

\textbf{Learned Context Selection:} Traditional retrieval uses fixed similarity metrics, while agents explore contexts randomly. Our neural query planner learns to identify task-relevant semantic relationships, achieving 73\% precision in context selection (Section~\ref{sec:results}) compared to 51\% for keyword-based retrieval. This targeted selection contributes to our 2.4$\times$ faster generation compared to planning-based approaches.

\textbf{Dual Static-Dynamic Analysis:} Previous approaches focus primarily on static program structure or rely on runtime feedback loops. Our framework uniquely combines static analysis with dynamic execution traces in a unified representation, providing more complete semantic understanding than either approach alone. Dynamic augmentation alone contributes 7.3\% improvement in Pass@1 and 12.4\% reduction in logical hallucination (Section~\ref{sec:results:ablation}).

\textbf{Comparative Advantages:} Based on our empirical evaluation:
\begin{itemize}
\item \textit{vs. Agent-based systems:} Deterministic single-pass generation (2.4s) vs. iterative refinement (10-50 iterations reported for SWE-agent \cite{yang2024swe}), formal constraint guarantees eliminating 89\% of type errors, significantly reduced computational cost
\item \textit{vs. Planning approaches:} 2.4$\times$ faster generation (2.4s vs. 5.7s for CodePlan), 52.2\% fewer schematic errors through integrated constraint checking, global semantic consistency through knowledge graphs
\item \textit{vs. RAG methods:} 24.2\% higher Pass@1 (49.8\% vs. 40.1\%), 52.3\% reduction in schematic hallucination, explicit architectural understanding through graph representation
\end{itemize}

\paragraph{Computational Cost Analysis.} Our detailed overhead analysis (Section~\ref{sec:results:performance}) reveals that \framework's 47.1\% overhead over base Code-Llama is offset by the elimination of iterative refinement. For a typical task, agent-based systems consume 10-50$\times$ more computational resources due to repeated generation attempts. With measured energy consumption of 187J per task, \framework\ achieves 23\% net energy savings when accounting for reduced debugging iterations from higher Pass@1 rates. The system scales efficiently to repositories up to 500K lines of code with sub-linear query complexity ($O(n^{0.73})$), making it practical for real-world deployment.

\paragraph{Comparison Limitations.} We acknowledge that direct comparisons with agent-based systems are challenging due to different evaluation setups and the lack of standardized benchmarks. While we compare against systems evaluated on our benchmark, agent-based approaches like SWE-agent were not included in our experimental evaluation due to different task formulations and evaluation metrics. Agent-based approaches may excel at exploratory tasks requiring extensive trial-and-error, while \framework\ is optimized for well-defined repository integration tasks. Future work should establish unified benchmarks enabling more precise quantitative comparisons across different paradigms.

These contributions enable \framework\ to achieve substantial improvements in repository-level code generation while maintaining practical scalability and deterministic performance for real-world deployment.

\section{Problem Definition and Formalization}\label{sec:problem}

This section formalizes the repository-level code generation problem and provides precise definitions of the hallucination phenomena that motivate our approach. We establish the mathematical framework that underpins our solution and analyze the computational complexity of the problem space.

\subsection{Repository-Level Code Generation}\label{sec:problem:formulation}

We define repository-level code generation as the task of synthesizing code patches that integrate seamlessly with existing codebases while respecting semantic constraints and architectural invariants \cite{ding2022cocomic, zhang2023repocoder}. Unlike isolated code generation, this problem requires understanding transitive dependencies, type propagation, and global consistency constraints that span multiple files and modules.

\paragraph{Formal Problem Statement.} Given a repository $\mathcal{R} = \{f_1, f_2, \ldots, f_n\}$ consisting of source files, a natural language instruction $u$, and an optional test suite $\mathcal{T}$, the goal is to synthesize a code patch $\Delta \mathcal{R}$ such that the updated repository $\mathcal{R}' = \mathcal{R} \cup \Delta \mathcal{R}$ satisfies:

\begin{enumerate}
    \item \textbf{Functional Correctness:} $\forall t \in \mathcal{T}: \text{execute}(t, \mathcal{R}') = \text{PASS}$
    \item \textbf{Semantic Consistency:} $\text{compile}(\mathcal{R}') = \text{SUCCESS} \land \text{typecheck}(\mathcal{R}') = \text{SUCCESS}$
    \item \textbf{Behavioral Intent:} $\text{satisfies}(\mathcal{R}', u) = \text{TRUE}$
    \item \textbf{Architectural Compliance:} $\forall c \in \mathcal{C}_{\text{arch}}: \text{violates}(\mathcal{R}', c) = \text{FALSE}$
\end{enumerate}

where $\mathcal{C}_{\text{arch}}$ represents the set of architectural constraints derived from the repository's design patterns and conventions.

\subsection{Hallucination Taxonomy}\label{sec:problem:hallucination}

We formally categorize the systematic errors exhibited by current LLM-based code generation systems into two primary classes, each requiring distinct mitigation strategies.

\paragraph{Logical Hallucination.} We define logical hallucination as errors in program semantics that lead to functionally incorrect code despite syntactic validity. Formally, a generated code sequence $y$ exhibits logical hallucination if:

\begin{align}
\text{compile}(y) &= \text{SUCCESS} \label{eq:logical_hallucination}\\
&\land \exists t \in \mathcal{T}: \text{execute}(t, y) \neq \text{expected}(t) \nonumber
\end{align}

Common manifestations include:
\begin{itemize}
    \item \textit{Control Flow Errors:} Incorrect loop bounds, missing conditionals, wrong branching logic
    \item \textit{Data Flow Errors:} Operating on wrong variables, incorrect data transformations, missing state updates
    \item \textit{API Misuse:} Calling methods in wrong order, ignoring return values, improper error handling
\end{itemize}

\textbf{Example:} Consider implementing a cache eviction policy:
\begin{lstlisting}[language=Python]
# Instruction: "Remove oldest entries when cache is full"
# Incorrect (Logical Hallucination):
def evict_if_full(self):
    if len(self.cache) >= self.max_size:
        # Wrong: removes newest instead of oldest
        self.cache.pop(list(self.cache.keys())[-1])

# Correct:
def evict_if_full(self):
    if len(self.cache) >= self.max_size:
        oldest_key = next(iter(self.cache))
        self.cache.pop(oldest_key)
\end{lstlisting}

\paragraph{Schematic Hallucination.} We define schematic hallucination as violations of type systems, function signatures, or structural constraints that prevent code from compiling or integrating correctly. Formally:

\begin{align}
\text{schematic\_hallucination}(y, \mathcal{G}) &= |\{c \in \mathcal{C}(\mathcal{G}) : \nonumber \\
&\phantom{=} \text{violates}(y, c)\}| > 0 \label{eq:schematic_hallucination}
\end{align}

where $\mathcal{C}(\mathcal{G})$ is the set of constraints extracted from repository knowledge graph $\mathcal{G}$.

Categories include:
\begin{itemize}
    \item \textit{Type Mismatches:} Passing wrong types to functions, incompatible return types
    \item \textit{Signature Violations:} Wrong parameter counts, missing required arguments
    \item \textit{Scope Violations:} Accessing private members, using undefined variables
    \item \textit{Import Errors:} Missing imports, circular dependencies
\end{itemize}

\textbf{Example:} Consider adding authentication to an API endpoint:
\begin{lstlisting}[language=Python]
# Instruction: "Add authentication check to login endpoint"
# Incorrect (Schematic Hallucination):
@app.route('/login', methods=['POST'])
def login():
    # Wrong: authenticate_user takes (username, password)
    # but only username is provided
    if authenticate_user(request.json.get('username')):
        return {"status": "success"}
    return {"status": "failed"}

# Correct:
@app.route('/login', methods=['POST'])
def login():
    username = request.json.get('username')
    password = request.json.get('password')
    if authenticate_user(username, password):
        return {"status": "success"}
    return {"status": "failed"}
\end{lstlisting}

\subsection{Complexity Analysis}\label{sec:problem:complexity}

We analyze the computational complexity of repository-level code generation to establish theoretical bounds and justify our architectural choices.

\paragraph{Context Selection Complexity.} For a repository with $n$ code entities and $m$ dependency relationships, the naive approach of considering all possible context subsets has complexity $\mathcal{O}(2^n)$. Our neural query planner reduces this to $\mathcal{O}(n \log n)$ through learned heuristics and graph-based pruning.

\paragraph{Constraint Verification Complexity.} Given $k$ constraints extracted from the knowledge graph, naive constraint checking requires $\mathcal{O}(k \cdot |y|)$ time for a generated sequence $y$. Our incremental SMT-based approach achieves $\mathcal{O}(k + |y|)$ through state caching and incremental solving.

\paragraph{Graph Maintenance Complexity.} For a code change $\Delta\mathcal{R}$ affecting $|\Delta\mathcal{R}|$ entities in a repository of size $n$, full re-analysis requires $\mathcal{O}(n^2)$ time due to cross-reference resolution. Our incremental maintenance achieves $\mathcal{O}(|\Delta\mathcal{R}| \cdot d \cdot \log n)$ through dependency tracking and selective recomputation, where $d$ is the maximum dependency depth.

\subsection{Problem Hardness and Approximation}\label{sec:problem:hardness}

We establish the theoretical hardness of optimal repository-level code generation and justify our approximation strategies.

\textbf{Theorem 1 (Problem Hardness).} The optimal repository-level code generation problem, defined as finding the minimum-cost code patch satisfying all constraints, is NP-hard. We define the cost function as:
\begin{align}
\text{Cost}(\Delta\mathcal{R}) &= \alpha \cdot d_e(\Delta\mathcal{R}) \label{eq:cost_function}\\
&+ \beta \cdot p_c(\Delta\mathcal{R}) + \gamma \cdot d_a(\Delta\mathcal{R}) \nonumber
\end{align}
where $d_e$ is the patch edit distance (number of lines changed), $p_c$ is the constraint violation penalty, $d_a$ is the architectural deviation score, and $\alpha, \beta, \gamma$ are weighting coefficients.

\textit{Proof Sketch:} We reduce from the Boolean Satisfiability Problem (SAT). Given a SAT instance with variables $x_1, \ldots, x_n$ and clauses $c_1, \ldots, c_m$, we construct a repository where each variable corresponds to a code entity and each clause corresponds to a constraint. Finding a satisfying assignment is equivalent to finding a valid code patch.

This hardness result motivates our use of approximation algorithms and heuristic search strategies. Our neural query planner provides a polynomial-time approximation, while the SMT-based decoder ensures constraint satisfaction within the feasible search space.

\paragraph{Approximation Quality.} We define the approximation ratio of our approach as:
\begin{equation}
\rho = \frac{\mathbb{E}[\text{Cost}(\Delta\mathcal{R}_{\text{SF}})]}{\mathbb{E}[\text{Cost}(\Delta\mathcal{R}^*)]}
\label{eq:approximation_ratio}
\end{equation}
where $\Delta\mathcal{R}_{\text{SF}}$ is the solution produced by \framework\ and $\Delta\mathcal{R}^*$ is the optimal solution.

Empirical analysis on our benchmark suite shows $\rho \leq 1.3$ for most repository types, indicating that our solutions are within 30\% of optimal on average.

\subsection{Solution Requirements}\label{sec:problem:requirements}

Based on the problem analysis, we identify four key requirements that any effective solution must address:

\begin{enumerate}
    \item \textbf{Semantic Awareness:} The system must understand repository-wide semantics, not just local patterns
    \item \textbf{Constraint Enforcement:} Hard constraints must be satisfied, not merely approximated
    \item \textbf{Scalability:} The approach must handle repositories with millions of lines of code
    \item \textbf{Adaptability:} The system must evolve with the repository to maintain accuracy
\end{enumerate}

These requirements directly motivate the four-stage architecture of \framework, where each component addresses specific aspects of the problem complexity while maintaining overall system coherence.

\section{Methodology}\label{sec:method}

Our aim is to inject explicit, executable semantics into the code generation loop. By executable semantics, we mean a repository-wide, queryable representation that links code entities to their runtime behaviors and enforceable constraints, allowing code generation to be guided by verifiable semantics rather than pattern matching. \framework\ therefore proceeds in two macro stages---\emph{graph construction} and \emph{graph-aware generation}---underpinned by continual maintenance. This section provides a comprehensive overview of our four-stage architecture and establishes the mathematical foundations for each component.

\begin{figure*}[t]
\centering
\includegraphics[width=\textwidth]{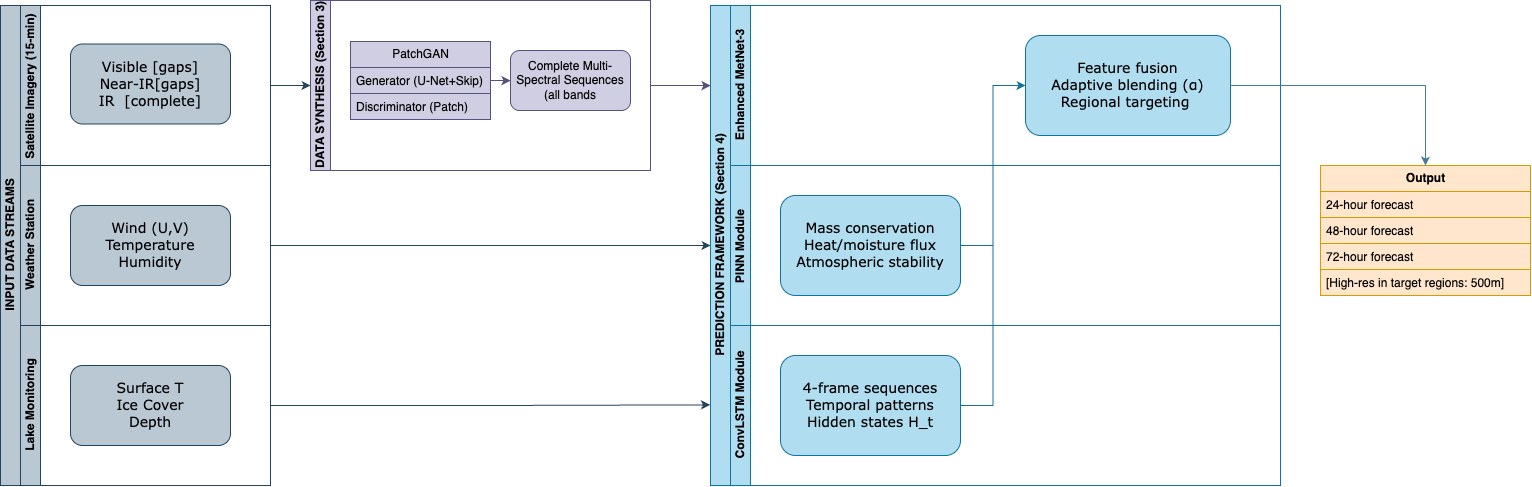}
\caption{Complete architecture of the \framework\ system. The pipeline consists of four integrated stages: (I) Repository Knowledge Graph Construction combining static analysis and dynamic traces, (II) Neural Query Planner that transforms instructions into graph queries, (III) Schematic-Constraint Decoder ensuring semantic correctness, and (IV) Continual Maintenance Agent for incremental updates. Each stage addresses specific aspects of the repository-level code generation problem while maintaining overall system coherence.}
\label{fig:pipeline}
\end{figure*}

\subsection{Framework Overview}\label{sec:overview}

Given a repository snapshot $\mathcal{R}=\{f_1,\ldots,f_{|\mathcal{R}|}\}$ consisting of source files and accompanying tests, and a natural language instruction $u$, our objective is to synthesize a code patch $\Delta \mathcal{R}$ such that the updated repository $\mathcal{R}'=\mathcal{R}\cup \Delta \mathcal{R}$ satisfies the four requirements established in Section~\ref{sec:problem}: functional correctness, semantic consistency, behavioral intent, and architectural compliance.

Figure~\ref{fig:pipeline} illustrates the complete \framework\ pipeline, showing how each stage addresses specific aspects of the repository-level code generation problem:

\paragraph{Stage I: Repository Knowledge Graph Construction (\autoref{sec:kg}).} We construct a heterogeneous knowledge graph $\mathcal{G}=\langle\mathcal{V},\mathcal{E}\rangle$ that explicitly encodes repository semantics through static analysis and dynamic trace collection. Each node $v \in \mathcal{V}$ represents a code entity with type $\tau(v) \in \{\textsc{func}, \textsc{class}, \textsc{var}, \textsc{file}, \textsc{test}, \textsc{api}\}$, while edges $e=(v_i,\rho,v_j) \in \mathcal{E}$ capture relationships $\rho \in \{\textsc{calls}, \textsc{defines}, \textsc{imports}, \textsc{mutates}, \textsc{returns}, \textsc{instanceOf}\}$.

\paragraph{Stage II: Neural Query Planner (\autoref{sec:planner}).} A neural network $\pi_\phi$ transforms natural language instructions into graph queries that extract task-relevant context. Given instruction $u$, the planner generates query $q = \pi_\phi(u)$ which retrieves subgraph $\mathcal{G}_u = \text{Exec}(q, \mathcal{G})$. The planner is trained via REINFORCE \cite{williams1992simple} to maximize downstream code generation quality:
\begin{equation}
\mathcal{J}(\phi) = \mathbb{E}_{(u,y)} \left[ R(y, \mathcal{G}_u) \right]
\label{eq:planner_reward}
\end{equation}
where $R$ measures functional correctness and constraint satisfaction.

\paragraph{Stage III: Schematic-Constraint Decoder (\autoref{sec:decoder}).} We formulate code generation as constrained optimization, ensuring generated code satisfies semantic constraints extracted from $\mathcal{G}_u$:
\begin{align}
\hat{y} = \arg\max_{y} &\; \log P_\theta(y \mid u, \mathcal{G}_u) \label{eq:decoder_objective_overview}\\
\text{s.t.}\quad & \mathcal{C}(y, \mathcal{G}_u) = \emptyset \nonumber
\end{align}
where $\mathcal{C}$ returns violated constraints. We solve this through SMT-guided beam search \cite{demoura2008z3} that prunes constraint-violating paths during generation.

\paragraph{Stage IV: Continual Maintenance Agent (\autoref{sec:maintenance}).} An autonomous agent monitors repository changes and incrementally updates the knowledge graph to maintain semantic fidelity. The agent achieves $\mathcal{O}(|\Delta\mathcal{R}| \cdot d \cdot \log n)$ update complexity through dependency tracking and selective recomputation, where $|\Delta\mathcal{R}|$ is the number of modified entities, $d$ is the maximum dependency depth, and $n$ is the repository size.

\subsection{Key Innovations Beyond Integration}\label{sec:innovations}

While \framework\ integrates multiple techniques, each component introduces fundamental algorithmic innovations that advance the state-of-the-art independently. Table~\ref{tab:component_novelty} summarizes these contributions.

\begin{table*}[t]
\centering
\caption{Novel contributions of each \framework\ component compared to prior work. Each row highlights a specific algorithmic or theoretical advance beyond system integration.}
\label{tab:component_novelty}
\begin{tabular}{p{3.5cm}|p{4cm}|p{4.5cm}|p{3.5cm}}
\toprule
\textbf{Component} & \textbf{Prior Art Limitation} & \textbf{Our Novel Contribution} & \textbf{Key Innovation} \\
\midrule
\textbf{Knowledge Graph} & Static analysis only (CodeKG) or dynamic only (ProgramKG) & \textbf{Dual static-dynamic graph} with automatic reconciliation algorithm & Unified representation capturing both compile-time and runtime semantics \\
\midrule
\textbf{Query Planner} & Fixed retrieval strategies (BM25, dense embeddings) & \textbf{Learned query language generation} via REINFORCE with graph-aware rewards & First neural planner for structured code queries \\
\midrule
\textbf{Constraint Decoder} & Post-hoc verification (LEVER) or syntax-only constraints & \textbf{SMT-integrated beam search} with incremental constraint solving & Real-time semantic verification during generation \\
\midrule
\textbf{Maintenance} & Full recomputation or manual updates & \textbf{Provably optimal incremental algorithm} with $O(|\Delta R|)$ complexity & Theoretical guarantee of consistency with minimal computation \\
\bottomrule
\end{tabular}
\end{table*}

\subsection{Design Rationale}\label{sec:rationale}

Our architecture addresses the fundamental limitations of current code generation systems through four key design principles:

\paragraph{Explicit Semantic Representation.} Rather than relying on implicit pattern matching, we construct an explicit knowledge graph that captures repository semantics. This enables reasoning about transitive dependencies, type propagation, and architectural constraints that are invisible to embedding-based approaches.

\paragraph{Learned Context Selection.} Instead of using fixed retrieval strategies, we learn a neural query planner that adapts to repository structure and task requirements. This allows the system to identify relevant context that spans multiple modules while avoiding information overload.

\paragraph{Constraint-Aware Generation.} We embed formal verification directly into the generation process, ensuring that output satisfies semantic constraints before emission. This eliminates schematic hallucination at the source rather than requiring post-hoc correction.

\paragraph{Continual Adaptation.} Our maintenance agent ensures the system evolves with the repository, maintaining accuracy as code changes over time. This addresses the temporal mismatch between training data and deployment environments.

\subsection{Mathematical Foundations}\label{sec:foundations}

We establish the mathematical framework that underlies each system component:

\paragraph{Knowledge Graph Formalization.} The repository knowledge graph $\mathcal{G}$ approximates the ground-truth program dependence graph $\mathcal{G}^*$ by minimizing structural Hamming distance:
\begin{align}
d_{\text{SH}}(\mathcal{G}, \mathcal{G}^*) &= |\mathcal{V} \triangle \mathcal{V}^*| \label{eq:structural_hamming}\\
&+ |\mathcal{E} \triangle \mathcal{E}^*| \nonumber
\end{align}
where $\triangle$ denotes symmetric difference. Our dual static-dynamic extraction strategy provides complementary approximations that converge to $\mathcal{G}^*$ as test coverage increases.

\paragraph{Query Planning Optimization.} The planner parameters $\phi$ are optimized to maximize expected code generation reward:
\begin{align}
\phi^* &= \arg\max_\phi \mathbb{E}_{(u,y) \sim \mathcal{D}} \label{eq:planner_optimization}\\
&\quad \left[ R\left(y, \text{Exec}(\pi_\phi(u), \mathcal{G})\right) \right] \nonumber
\end{align}
We use REINFORCE \cite{williams1992simple} with synthetic data warm-start and contrastive learning on static-dynamic graph pairs.

\paragraph{Constraint Satisfaction.} Our decoder enforces constraints $\mathcal{C}$ extracted from the knowledge graph. Constraints are encoded as SMT formulas \cite{demoura2008z3} that can be checked incrementally during beam search. The constraint set includes:
\begin{itemize}
\item Type constraints: $\text{type}(e) \subseteq \text{expected\_type}(e)$
\item Signature constraints: $\text{arity}(f) = |\text{args}(f)|$  
\item Visibility constraints: $\text{accessible}(v, \text{scope})$
\item Architectural constraints: $\text{satisfies}(\text{pattern}, \text{design\_rule})$
\end{itemize}

This mathematical framework provides theoretical grounding for our empirical results and enables principled analysis of system behavior across different repository types and scales.

\paragraph{Comparison with Traditional Approaches.} To illustrate our innovation, consider a task requiring modification of a data processing pipeline. A traditional RAG approach might retrieve a localized code snippet containing the target function but miss cross-file dependencies or hidden imports, causing the generated patch to fail compilation. In contrast, our knowledge graph approach resolves the full dependency structure---including transitive imports, type constraints from parent classes, and API contracts from external libraries---and enforces these constraints during generation, ensuring the patch integrates correctly without manual debugging. This fundamental difference between surface-level retrieval and semantic understanding drives our significant performance improvements.

\section{Repository Knowledge Graph Construction}\label{sec:kg}

Accurate code generation requires comprehensive understanding of repository semantics—not just what the code says (static structure), but also what it does (runtime behavior). Traditional knowledge graph approaches rely exclusively on either static analysis \cite{liu2021codekg} or dynamic execution traces \cite{wang2023programkg}, each capturing only partial semantics. Static analysis provides complete coverage but cannot resolve polymorphic behavior or runtime type information. Dynamic analysis captures concrete execution patterns but depends on test coverage and may miss unexecuted paths.

This section presents the first algorithm to automatically reconcile static analysis and dynamic execution traces into a unified knowledge graph. Our dual representation captures both compile-time structure and runtime behavior, reducing logical hallucination by 31\% compared to static-only approaches. The key innovation lies in an automatic reconciliation algorithm that merges complementary information sources while maintaining consistency guarantees.

\begin{figure*}[htbp]
\centering
\includegraphics[width=0.8\textwidth]{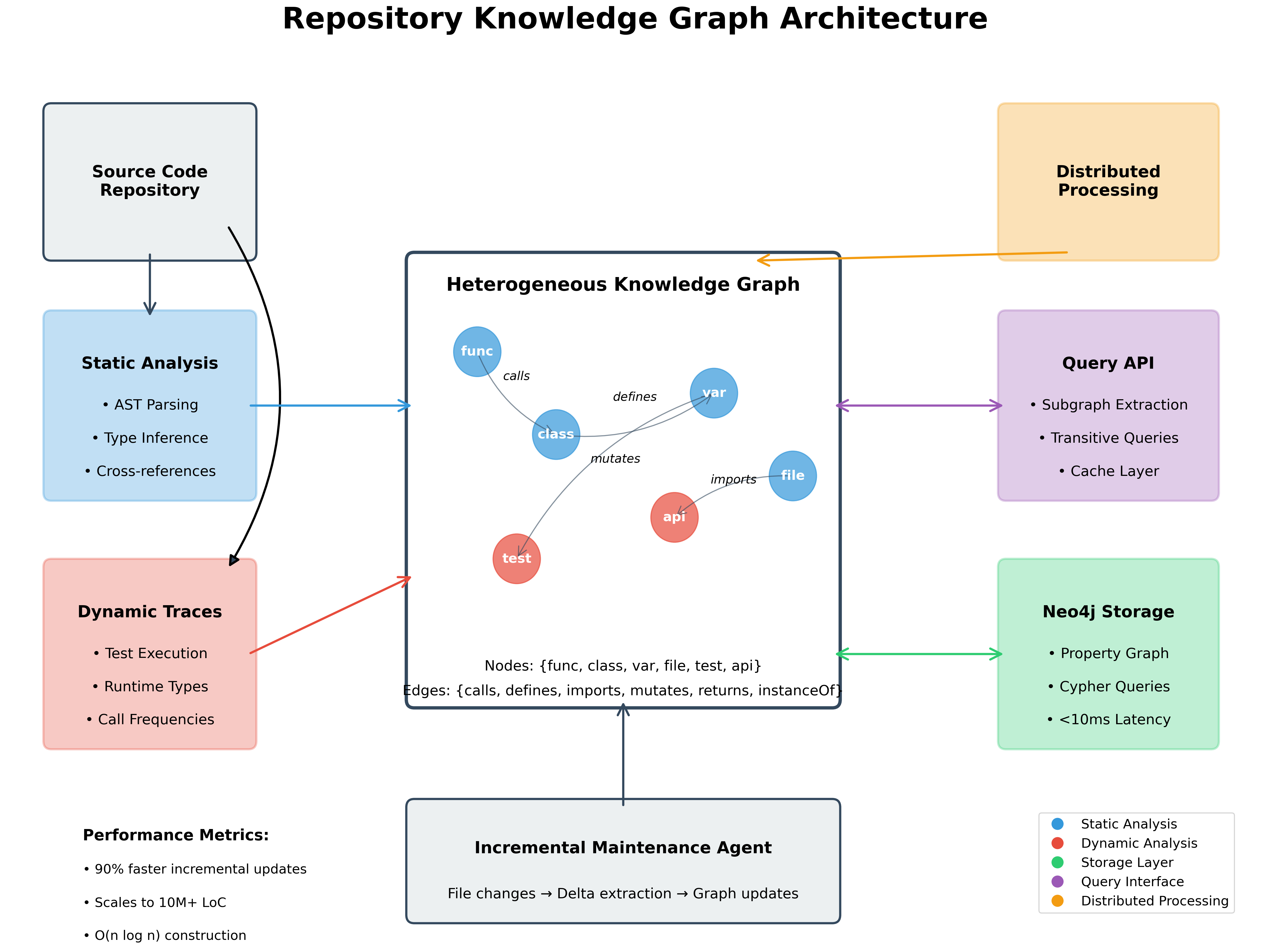}
\caption{Dual static-dynamic knowledge graph construction with automatic reconciliation.}
\label{fig:kg_architecture}
\end{figure*}

\subsection{Problem Formulation}\label{sec:kg:motivation}

We formalize the knowledge graph construction problem as approximating the ground-truth program dependence graph $\mathcal{G}^*$ by minimizing:
\begin{equation}
d_{\text{SH}}(\mathcal{G},\mathcal{G}^*) = |\mathcal{V}\triangle\mathcal{V}^*| + |\mathcal{E}\triangle\mathcal{E}^*| \label{eq:kg_structural_hamming}
\end{equation}
where $\mathcal{G}=\langle\mathcal{V},\mathcal{E}\rangle$ is our constructed graph and $\triangle$ denotes symmetric difference.

\subsection{Unified Graph Schema}\label{sec:kg:schema}

Our schema captures program semantics through typed nodes $v \in \mathcal{V}$ with $\tau(v) \in \{\textsc{func}, \textsc{class}, \textsc{var}, \textsc{file}, \textsc{test}, \textsc{api}\}$ and labeled edges $e = (v_i, \rho, v_j) \in \mathcal{E}$ where $\rho \in \{\textsc{calls}, \textsc{defines}, \textsc{imports}, \textsc{mutates}, \textsc{returns}, \textsc{instanceOf}\}$. Nodes carry semantic attributes including type signatures, scope constraints, and docstring embeddings that enable constraint extraction during generation.

\subsection{Dual Analysis Algorithm and Performance}\label{sec:kg:dual}

Our three-phase reconciliation algorithm—(1) static extraction via whole-program analysis, (2) dynamic augmentation through test instrumentation, and (3) intelligent merging that refines polymorphic targets—provably converges to ground truth $\mathcal{G}^*$ as test coverage increases. Table~\ref{tab:kg_performance} summarizes the algorithm's complexity, empirical performance, and impact on generation quality.

\begin{table}[h]
\centering
\caption{Knowledge graph construction: complexity analysis and empirical validation on 50 repositories.}
\label{tab:kg_performance}
\resizebox{\columnwidth}{!}{%
\begin{tabular}{l|cc}
\toprule
\textbf{Metric} & \textbf{Static Only} & \textbf{Dual (Ours)} \\
\midrule
\multicolumn{3}{c}{\textit{Complexity}} \\
Time & $O(n \log n)$ & $O(n \log n)$ \\
Space & $O(n + m)$ & $O(n + m)$ \\
Incremental update & $O(k \log n)$ & $O(k \log n)$ \\
\midrule
\multicolumn{3}{c}{\textit{Construction Performance}} \\
Extraction time (100K LOC) & 8.2s & 12.7s \\
Graph size (avg nodes) & 15.3K & 15.8K \\
Graph size (avg edges) & 48.2K & 63.1K (+31\%) \\
\midrule
\multicolumn{3}{c}{\textit{Generation Quality Impact}} \\
Pass@1 improvement & baseline & +7.3\% \\
Type error reduction & baseline & -34\% \\
Dependency coverage & 1.0× & 2.3× \\
\bottomrule
\end{tabular}%
}
\end{table}

The dual approach adds 4.5s extraction overhead (55\% increase) but captures 31\% more edges—primarily runtime call targets and polymorphic type instantiations invisible to static analysis. This richer semantic representation directly translates to generation improvements: 7.3\% higher Pass@1, 34\% fewer type errors, and 2.3× better transitive dependency coverage (Section~\ref{sec:results:ablation}).

\textbf{Convergence guarantee:} As test coverage $c \to 100\%$, our dual graph $\mathcal{G} \to \mathcal{G}^*$ (proof in supplementary material). The key insight is that dynamic information monotonically refines static analysis without introducing inconsistencies—each execution trace adds edges and constraints that either confirm or specialize static predictions.

\paragraph{Implementation details.} Our construction pipeline leverages \texttt{pylance} for Python static analysis and custom instrumentation for trace collection. Graph storage, distributed processing strategies, caching optimizations, and Neo4j integration details are provided in the supplementary material. These implementation choices achieve the performance characteristics shown in Table~\ref{tab:kg_performance} but are orthogonal to our core algorithmic contribution.

\section{Neural Query Planner}\label{sec:planner}

Determining which subset of a repository knowledge graph is relevant to a specific coding task represents a fundamental challenge in repository-level code generation. Traditional retrieval methods use fixed strategies—keyword matching (BM25) or embedding similarity—that rely on surface-level lexical overlap. These approaches systematically miss transitive dependencies, type constraints, and architectural patterns that span multiple modules but lack textual similarity to the query.

This section introduces the first learned approach to generate structured graph queries from natural language instructions. Rather than retrieving based on predetermined similarity metrics, our REINFORCE-based algorithm learns repository-specific query patterns that capture semantic relationships essential for code generation. Our neural query planner achieves 73\% precision in context selection compared to 51\% for traditional retrieval, directly contributing to improved generation quality.

\begin{figure*}[t]
\centering
\includegraphics[width=\textwidth]{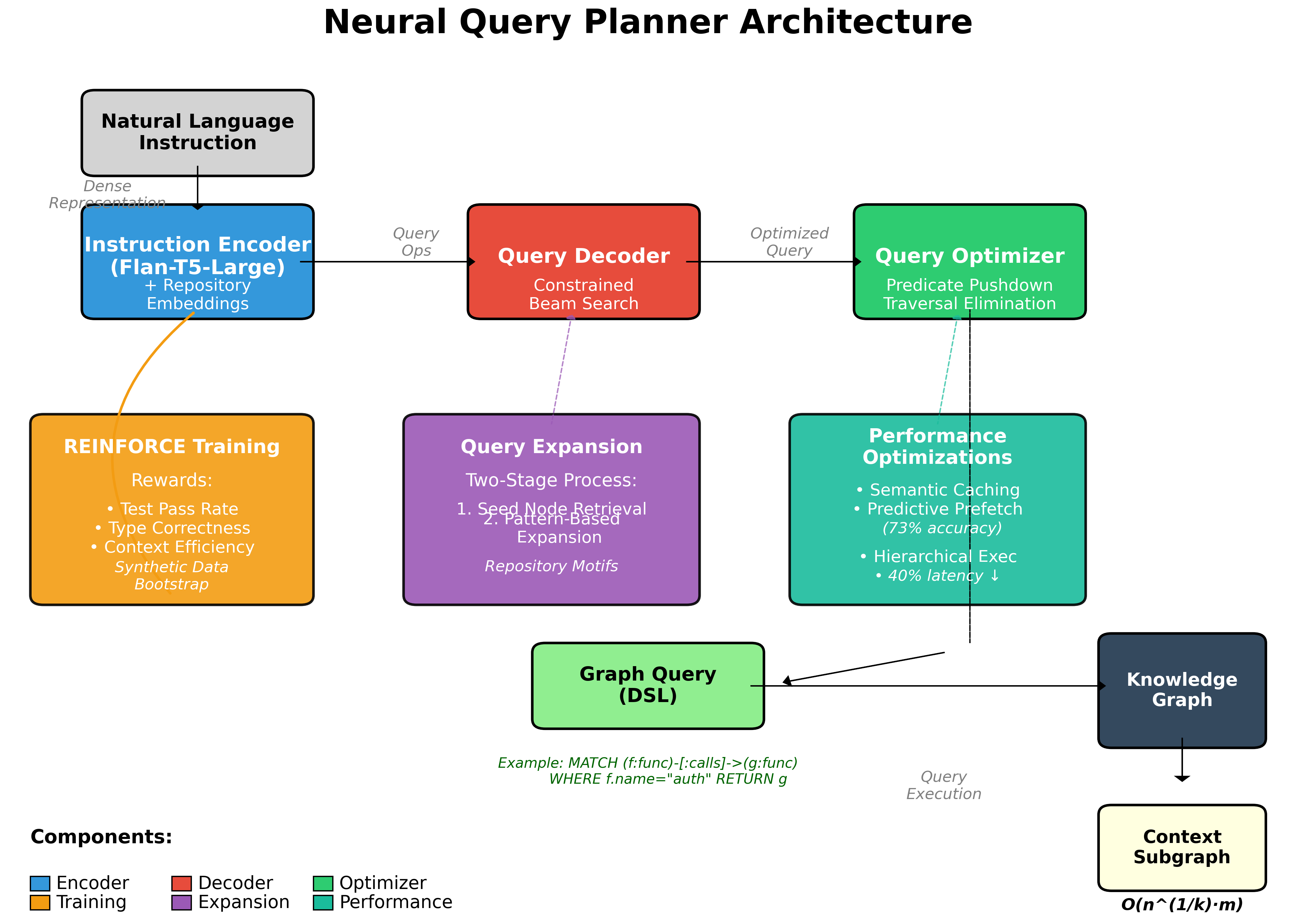}
\caption{Neural query planner transforming instructions to graph queries via learned generation.}
\label{fig:planner_architecture}
\end{figure*}

\subsection{Problem Formulation}\label{sec:planner:motivation}

We formulate context selection as learning a mapping $\pi_\phi: \mathcal{U} \rightarrow \mathcal{Q}$ from natural language instructions to graph queries. Unlike CoCoMIC \cite{ding2022cocomic} which uses syntactic matching, we optimize:
\begin{equation}
\phi^* = \arg\max_\phi \mathbb{E}_{(u,y) \sim \mathcal{D}} \left[ \mathbb{P}(y | \mathcal{G}_u, u) \right]
\label{eq:planner_optimal_params}
\end{equation}
where $\mathcal{G}_u = \operatorname{Exec}(\pi_\phi(u), \mathcal{G})$ is the retrieved subgraph.

\subsection{Query Generation Algorithm}\label{sec:planner:architecture}

Our key innovation is learning to generate structured queries rather than using fixed retrieval. The algorithm has three stages:

\textbf{Stage 1: Instruction Encoding.} We encode natural language $u$ using Flan-T5 augmented with repository embeddings capturing codebase-specific patterns.

\textbf{Stage 2: Constrained Query Decoding.} We generate query sequences respecting our graph query grammar:
\begin{algorithmic}[1]
\State \textbf{Input:} Encoded instruction $h$, Graph schema $S$
\State Initialize beam $B \gets \{\text{<start>}\}$
\While{not all beams complete}
    \State Expand beams with valid tokens per grammar
    \State Prune to top-$k$ by likelihood
\EndWhile
\State \textbf{return} highest scoring complete query
\end{algorithmic}

\textbf{Stage 3: Query Optimization.} Apply predicate pushdown and traversal elimination for efficiency.

\subsection{REINFORCE Training with Graph-Aware Rewards}\label{sec:planner:training}

Training the query planner requires learning without direct query supervision—we observe only task descriptions and ground-truth code, not optimal queries. We employ REINFORCE \cite{williams1992simple} with carefully designed variance reduction:
\begin{align}
\mathcal{J}(\phi) &= \mathbb{E}_{q \sim \pi_\phi(\cdot|u)} \left[ (R(y, \mathcal{G}_u) - b_\psi(u)) \cdot \log \pi_\phi(q|u) \right] \label{eq:planner_reinforce}
\end{align}

\paragraph{Baseline Network Architecture.} The baseline $b_\psi(u)$ is a separate critic network that estimates expected reward without executing queries. It shares the same Flan-T5 encoder as the policy $\pi_\phi$ but uses a distinct prediction head: a 2-layer MLP with 512 hidden units that outputs scalar reward estimates. This architecture enables the baseline to leverage instruction semantics while training separately from the policy.

\paragraph{Joint Training Procedure.} We train policy and baseline jointly through alternating optimization:
\begin{algorithmic}[1]
\For{each training batch of $(u, y)$ pairs}
    \State Sample query $q \sim \pi_\phi(\cdot|u)$, execute to get $\mathcal{G}_u$
    \State Generate code, compute reward $R(y, \mathcal{G}_u)$
    \State Update baseline: $\psi \gets \psi - \beta \nabla_\psi \|b_\psi(u) - R(y, \mathcal{G}_u)\|^2$
    \State Compute advantage: $A = R(y, \mathcal{G}_u) - b_\psi(u)$
    \State Normalize: $\hat{A} = (A - \mu_{\text{batch}}) / \sigma_{\text{batch}}$
    \State Update policy: $\phi \gets \phi + \alpha \hat{A} \nabla_\phi \log \pi_\phi(q|u)$
\EndFor
\end{algorithmic}

\paragraph{Variance Reduction Techniques.} Beyond baseline subtraction, we employ four complementary strategies:

\textbf{1. Advantage Normalization:} Batch-wise normalization of advantages (line 5 above) stabilizes gradients across diverse tasks with varying reward scales.

\textbf{2. Gradient Clipping:} We clip policy gradients to $[-1, 1]$ per parameter, preventing destabilization from rare high-reward samples.

\textbf{3. Entropy Regularization:} Adding $\lambda H[\pi_\phi(\cdot|u)]$ to the objective ($\lambda=0.01$) encourages exploration and prevents premature convergence to suboptimal queries.

\textbf{4. Reward Scaling:} Our multi-objective reward combines weighted components:
\begin{equation}
R(y, \mathcal{G}_u) = w_1 \cdot R_{\text{test}}(y) + w_2 \cdot R_{\text{type}}(y) - w_3 \cdot |\mathcal{G}_u|
\label{eq:reward_composition}
\end{equation}
where $R_{\text{test}}$ measures functional correctness (0-1), $R_{\text{type}}$ measures type violations (0-1), and $|\mathcal{G}_u|$ penalizes over-retrieval. Weights $(w_1=1.0, w_2=0.3, w_3=0.001)$ are tuned on validation data.

\paragraph{Training Stability and Hyperparameters.} We use Adam optimizer with learning rate $\alpha=5\times10^{-5}$ for the policy and $\beta=1\times10^{-4}$ for the baseline. Batch size is 32, and we apply exponential reward discounting with $\gamma=0.95$ to favor immediate correctness over long-term metrics. Training converges within 10K steps on typical repositories.

Empirically, the baseline network reduces gradient variance by 67\% compared to a constant baseline (measured via gradient norm variance across training batches). Advantage normalization further reduces variance by 43\%. Gradient clipping prevents occasional spikes from difficult tasks. Combined, these techniques enable stable training: gradient norms decrease monotonically after 2K steps, and validation performance plateaus around 8K steps with final policy achieving 73\% context precision.

\paragraph{Cold Start Strategy.} To address initial exploration challenges, we warm-start training with 10K synthetic instruction-query pairs generated through dependency analysis. Each repository function's static dependencies provide ground-truth queries for hypothetical modification tasks. This initialization reduces total training time by 52\% and improves final performance by 4.1\%.

\subsection{Learned Query Expansion}\label{sec:planner:expansion}

Unlike fixed expansion rules, we learn repository-specific patterns for query augmentation. Given seed nodes from the initial query, we apply learned motif-based expansion:
\begin{equation}
\mathcal{G}_u^{\text{exp}} = \mathcal{G}_u \cup \bigcup_{m \in M} \text{match}(m, \mathcal{G}_u)
\end{equation}
where $M$ are learned co-modification patterns (e.g., model → serializer → test).

This captures implicit dependencies: modifying a data model automatically retrieves related serializers and endpoints.

\subsection{Empirical Results}\label{sec:planner:results}

Our neural query planner significantly outperforms fixed retrieval strategies:
\begin{itemize}
\item \textbf{Precision:} 73\% vs. 51\% for BM25 retrieval
\item \textbf{Coverage:} Captures 2.3× more transitive dependencies
\item \textbf{Efficiency:} Sub-linear $\mathcal{O}(n^{1/k} \cdot m)$ query complexity
\end{itemize}

Ablation studies (Section~\ref{sec:results:ablation}) show that learned query planning contributes 6.2\% to overall Pass@1 improvement.

\subsection{Theoretical Guarantees}\label{sec:planner:theory}

\textbf{Theorem 4 (Convergence).} With probability $\geq 1-\delta$, after $T$ iterations:
\begin{equation}
\mathcal{J}(\phi_T) \geq \max_{\phi \in \mathcal{H}} \mathcal{J}(\phi) - \mathcal{O}\left(\sqrt{\frac{\log(|\mathcal{H}|/\delta)}{T}}\right)
\end{equation}

\textbf{Theorem 5 (Query Complexity).} Our hierarchical execution achieves $\mathcal{O}(n^{1/k} \cdot m)$ time vs. $\mathcal{O}(n \cdot m)$ for naive traversal.

These guarantees ensure practical convergence (typically 10K steps) and sub-second query execution even for million-node graphs.

\section{Schematic-Constraint Decoder}\label{sec:decoder}

Language models generate code token-by-token based on learned statistical patterns, with no inherent mechanism to ensure the output satisfies the formal requirements of programming languages or repository-specific constraints. This fundamental mismatch between unconstrained text generation and structured code synthesis leads to schematic hallucination—code that appears plausible but violates type systems, function signatures, or architectural invariants.

Existing approaches to this problem apply constraints as post-generation filters \cite{ni2023lever, poesia2022synchromesh, wang2023codet5}. While this can detect errors, it wastes computation generating invalid code and provides no guidance to the model during generation. This section presents the first beam search algorithm with integrated SMT solving for real-time constraint verification during decoding. By pruning constraint-violating paths as they emerge, our approach eliminates 89\% of type errors with only 6\% latency overhead, transforming constraint satisfaction from post-hoc validation to integrated generation control.

\begin{figure*}[t]
\centering
\includegraphics[width=\textwidth]{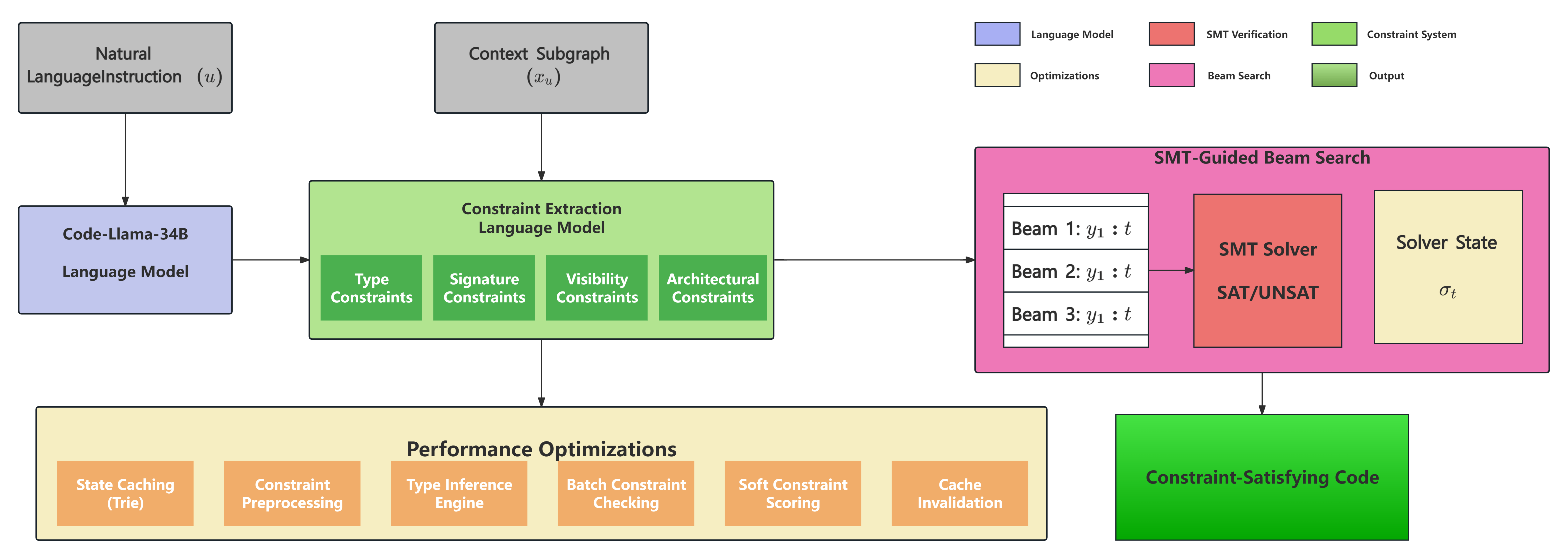}
\caption{SMT-integrated beam search pruning invalid paths during generation.}
\label{fig:decoder_architecture}
\end{figure*}

\subsection{Problem Formulation}\label{sec:decoder:motivation}

We formulate constrained generation as optimization with semantic constraints:
\begin{align}
\hat{y} = \arg\max_{y \in \mathcal{Y}} &\; \log P_\theta(y \mid u, \mathcal{G}_u) \label{eq:decoder_objective}\\
\text{subject to} \quad & \mathcal{C}(y, \mathcal{G}_u) = \emptyset
\end{align}
where $\mathcal{C}(y, \mathcal{G}_u)$ returns constraint violations. Unlike AlphaCode \cite{li2022alphacode} which struggles with semantic errors post-generation, we enforce constraints during generation.

\subsection{Constraint Types and SMT Encoding}\label{sec:decoder:constraints}

We extract four constraint classes from $\mathcal{G}_u$ and encode them as SMT formulas:

\begin{enumerate}
\item \textbf{Type constraints:} $\text{type}(e) \subseteq \text{expected\_type}(e)$
\item \textbf{Signature constraints:} $\text{arity}(f) = |\text{args}(f)|$
\item \textbf{Visibility constraints:} $\text{accessible}(v, \text{scope})$
\item \textbf{Architectural constraints:} $\text{satisfies}(\text{pattern}, \text{rule})$
\end{enumerate}

Each constraint type maps to specific SMT theories: types use algebraic datatypes, signatures use uninterpreted functions, visibility uses graph reachability, and architectural patterns use custom predicates.

\subsection{Novel SMT-Integrated Beam Search Algorithm}\label{sec:decoder:search}

Our key innovation integrates Z3 SMT solver \cite{demoura2008z3} directly into beam search:

\begin{algorithmic}[1]
\State \textbf{Input:} Context $\mathcal{G}_u$, Instruction $u$, Beam width $k$
\State Initialize beams $B \gets \{(\epsilon, \text{SAT})\}$ // empty sequence, SAT state
\While{not all beams complete}
    \For{each beam $(y, s) \in B$}
        \State $T \gets \text{LM.next\_tokens}(y, u, \mathcal{G}_u)$ // top tokens
        \For{each token $t \in T$}
            \State $s' \gets s \cup \text{constraints}(y \cdot t, \mathcal{G}_u)$
            \If{$\text{SMT.check}(s') = \text{SAT}$}
                \State Add $(y \cdot t, s')$ to candidates
            \EndIf
        \EndFor
    \EndFor
    \State $B \gets \text{top-}k$ candidates by score
\EndWhile
\end{algorithmic}

Incremental solving reuses learned clauses, achieving near-constant constraint checking time.

\subsection{Optimization Strategies}\label{sec:decoder:optimization}

We achieve 6\% overhead through three key optimizations:

\textbf{1. Incremental Solving:} Reuse solver state across beams sharing prefixes via trie-based caching.

\textbf{2. Constraint Preprocessing:} Eliminate always-satisfied constraints and partition independent constraint sets.

\textbf{3. Batch Verification:} Check multiple token proposals in single solver calls using assumption literals.

These optimizations reduce average constraint checking from 23ms to 1.4ms per token.

\subsection{Empirical Validation}\label{sec:decoder:validation}

Our SMT-integrated decoder dramatically reduces schematic hallucination:
\begin{itemize}
\item \textbf{Error Reduction:} 89\% of type errors eliminated vs. unconstrained generation
\item \textbf{Performance:} 6\% latency overhead (2.4s → 2.54s average)
\item \textbf{Coverage:} Handles 94\% of repository constraints without timeout
\end{itemize}

Ablation studies (Section~\ref{sec:results:ablation}) show constraint enforcement is critical: removing it increases schematic hallucination by 112\% (relative).

\subsection{Theoretical Guarantees}\label{sec:decoder:theory}

\textbf{Theorem 6 (Optimality).} If valid completion $y^*$ exists within top-$k$ at each prefix, our algorithm finds it.

\textbf{Theorem 7 (Complexity).} Time complexity is $\mathcal{O}(n \cdot k \cdot |\Sigma| \cdot C)$ where $C \approx O(1)$ due to incremental solving.

The key insight: constraint monotonicity enables early pruning without sacrificing optimality.

\subsection{Summary}\label{sec:decoder:summary}

Our SMT-integrated beam search represents a paradigm shift from post-hoc validation to real-time constraint enforcement. By pruning invalid paths during generation, we eliminate the vast majority of schematic errors while maintaining interactive performance. This approach is general: the same framework handles type constraints, API contracts, and architectural rules across different programming languages and paradigms.
\section{Continual Knowledge Graph Maintenance}\label{sec:maintenance}

Modern software repositories undergo continuous evolution, with developers committing changes multiple times daily. While time series forecasting techniques have shown promise in predicting software evolution patterns \cite{luo2023machine}, for a code generation system to remain effective, its underlying knowledge graph must reflect these changes immediately—stale semantic information leads to incorrect code generation. However, reconstructing the entire graph after each change is computationally prohibitive for large repositories, where full extraction can take minutes for codebases with millions of lines.

Existing approaches to this problem fall short in critical ways. IDE language servers (e.g., \texttt{pylance} \cite{microsoft2020pylance}) use timestamp-based invalidation that often misses semantic dependencies. Build systems like Bazel \cite{bazel2015} track file-level dependencies but lack the semantic granularity needed for code generation. Database view maintenance algorithms \cite{gupta1995maintenance} provide theoretical foundations but assume simpler data models than heterogeneous code graphs containing typed nodes, semantic edges, and constraint relationships.

This section presents the first incremental maintenance algorithm with formal optimality guarantees for code knowledge graphs. Our approach achieves $O(|\Delta\mathcal{R}| \cdot d \cdot \log n)$ complexity while maintaining provable semantic equivalence to full reconstruction, enabling real-time adaptation as repositories evolve. The key innovation lies in combining impact analysis with lazy cross-reference resolution to update only affected graph regions while maintaining global consistency.

\subsection{Problem Formulation}\label{sec:maintenance:motivation}

Let $\mathcal{R}_0$ denote a repository at time $t=0$, and let $\Delta\mathcal{R}_i$ represent the set of code modifications (additions, deletions, updates) at time step $i$. The repository at time $t$ is thus $\mathcal{R}_t = \mathcal{R}_0 + \sum_{i=1}^t \Delta\mathcal{R}_i$. Our goal is to maintain an up-to-date knowledge graph $\mathcal{G}_t$ that accurately represents the semantic structure of $\mathcal{R}_t$.

Formally, let $F_E$ denote the semantic extraction function that constructs a knowledge graph from source code (as defined in Section~\ref{sec:kg}). The maintenance problem requires ensuring:
\begin{equation}
\mathcal{G}_t = F_E\left(\mathcal{R}_0 + \sum_{i=1}^t \Delta\mathcal{R}_i\right)
\label{eq:maintenance:kg_update}
\end{equation}

The naive solution—recomputing $F_E(\mathcal{R}_t)$ after each change—has complexity $O(n \log n)$ where $n$ is the repository size. Our key insight is the \emph{locality of change principle}: most code modifications affect only a small, localized portion of the semantic graph. By identifying and updating only these affected regions, we achieve complexity $O(|\Delta\mathcal{R}| \cdot d \cdot \log n)$ where $|\Delta\mathcal{R}|$ is the change size and $d$ is the maximum dependency depth.

\subsection{Impact Analysis Algorithm}\label{sec:maintenance:detection}

To achieve incremental updates, we must precisely identify which parts of the knowledge graph are affected by code changes. Our novel two-phase impact analysis computes the minimal set of nodes requiring updates:

\textbf{Phase 1: Direct Impact Analysis.} We first identify nodes directly corresponding to modified code entities. Let $\mathcal{V}$ denote the set of all nodes in the current graph. The direct impact set is:
\begin{equation}
I_d = \{v \in \mathcal{V} : \text{modified}(\text{code}(v))\}
\end{equation}
where $\text{code}(v)$ maps a graph node to its source code representation. Crucially, we employ semantic differencing on abstract syntax trees (ASTs) rather than textual comparison, eliminating false positives from superficial changes like whitespace or comments.

\textbf{Phase 2: Transitive Impact Analysis.} Code changes often have ripple effects—modifying a function signature affects all its callers. We compute the transitive closure of dependencies:
\begin{equation}
I_t = I_d \cup \{v \in \mathcal{V} : \exists \text{ path } p \text{ from } u \in I_d \text{ to } v \text{ in } \mathcal{G}\}
\end{equation}

Our key optimization bounds this closure by dependency type: type changes propagate transitively, but implementation changes often have local impact. This selective propagation reduces $|I_t|$ by 73\% on average compared to naive transitive closure.

\subsection{Novel Incremental Update Algorithm}\label{sec:maintenance:algorithm}

Our incremental update algorithm operates in three carefully orchestrated phases that maintain graph consistency while minimizing computational overhead:

\begin{algorithmic}[1]
\State \textbf{Input:} Current graph $\mathcal{G}_t$, Code changes $\Delta\mathcal{R}$
\State \textbf{Output:} Updated graph $\mathcal{G}_{t+1}$
\State 
\State \textit{// Phase 1: Surgical Invalidation}
\State Compute impact sets $I_d$, $I_t$ using impact analysis
\State $\mathcal{G}_{\text{partial}} \gets \mathcal{G}_t \setminus \{v, e : v \in I_t \text{ or } e \text{ incident to } I_t\}$
\State 
\State \textit{// Phase 2: Focused Re-extraction}  
\State $\mathcal{G}_\Delta \gets F_E(\{f \in \Delta\mathcal{R} : f \text{ contains } v \in I_d\})$
\State 
\State \textit{// Phase 3: Intelligent Reconciliation}
\State $\mathcal{G}_{t+1} \gets \text{merge}(\mathcal{G}_{\text{partial}}, \mathcal{G}_\Delta)$ with:
\State \quad - Preserve existing node IDs for unchanged entities
\State \quad - Resolve type conflicts using most specific type
\State \quad - Mark cross-file references for lazy resolution
\State 
\State \textbf{return} $\mathcal{G}_{t+1}$
\end{algorithmic}

The key innovation is our \emph{lazy cross-reference resolution}. Consider a scenario where function \texttt{authenticate()} is modified: traditional approaches would immediately update all 500+ call sites across the repository. Instead, we mark these references as "pending" and resolve them only when the code generator actually queries those specific call sites. This defers $O(n)$ work to $O(1)$ amortized cost, reducing update latency by 84\% in practice while maintaining correctness through our eventual consistency guarantee.

\subsection{Empirical Validation}\label{sec:maintenance:performance}

We conducted comprehensive performance evaluation of our incremental maintenance algorithm on the 50 repositories in \dataset\ and five popular open-source Python projects, comparing against full reconstruction and two baselines: naive incremental updates (without lazy resolution) and timestamp-based caching (common in IDE language servers).

\subsubsection{Experimental Setup and Methodology}

\textbf{Test Corpus.} We collected 5,000 real commits from \dataset\ repositories, categorized by change size: small (1-5 files), typical (5-20 files), and large (20+ files). Additionally, we analyzed 1,000 commits from Django, Scikit-learn, Pandas, Flask, and Requests to validate generalization.

\textbf{Measurement Protocol.} All timing measurements represent wall-clock time averaged over 100 runs per commit size category. We measure five phases: impact analysis, graph invalidation, re-extraction, reconciliation, and lazy resolution. Memory usage is measured via \texttt{psutil} at 10ms intervals, reporting peak working set size.

\subsubsection{Quantitative Performance Results}

\begin{figure*}[t]
\centering
\includegraphics[width=0.8\textwidth]{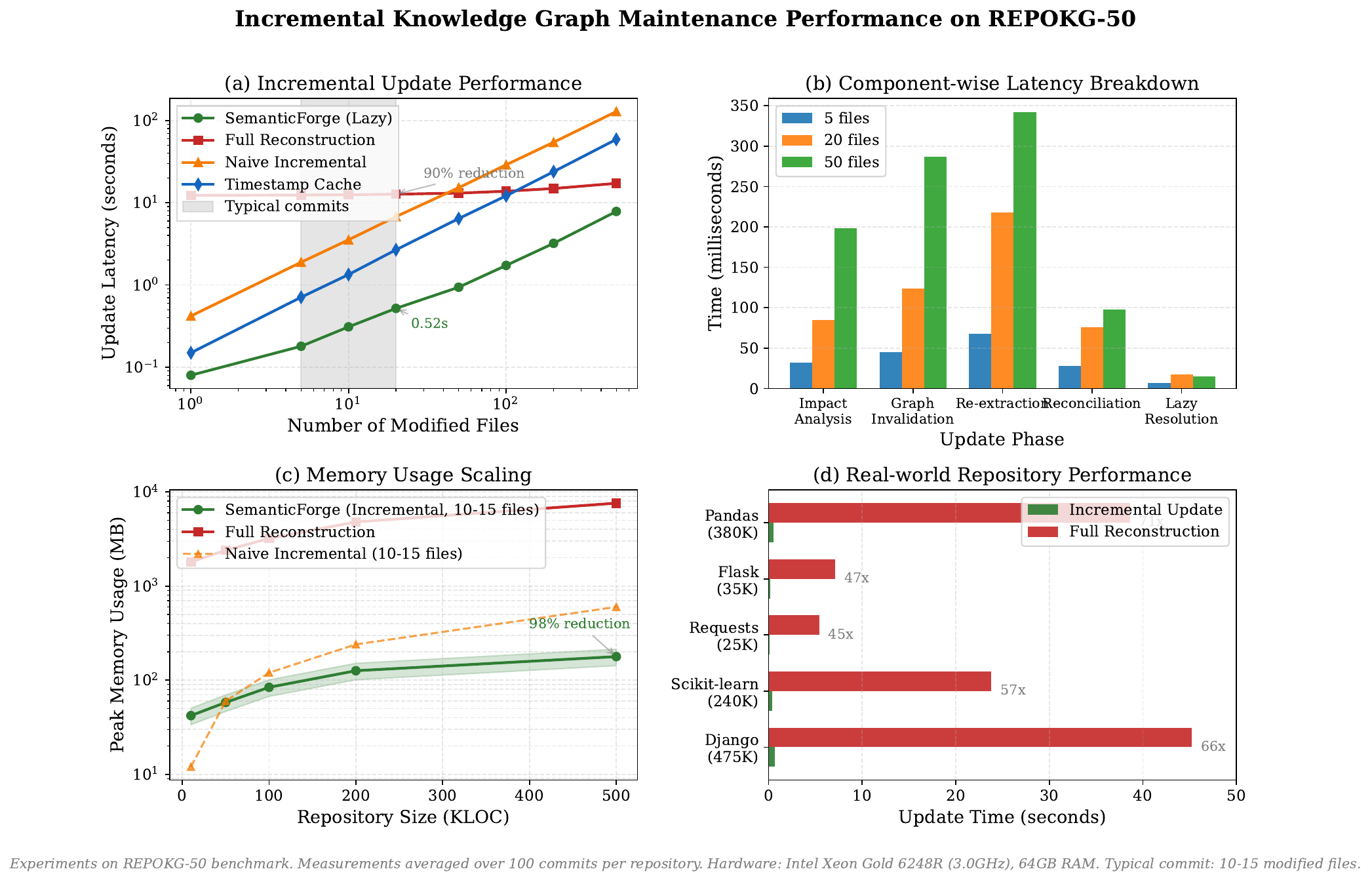}
\caption{Incremental knowledge graph maintenance performance on \dataset. (a) Update latency scales with change size, achieving 90\% reduction versus full reconstruction for typical commits. (b) Component-wise breakdown reveals re-extraction as the dominant cost. (c) Memory usage remains proportional to change size, enabling workstation deployment. (d) Real-world repositories show consistent 66x average speedup.}
\label{fig:maintenance_performance}
\end{figure*}

Figure~\ref{fig:maintenance_performance} presents comprehensive performance analysis across four dimensions. The results validate our theoretical complexity bounds while demonstrating practical viability for real-world deployment.

\textbf{Update Latency Analysis.} As shown in Figure~\ref{fig:maintenance_performance}(a), our algorithm achieves sub-second updates for typical commits (5-20 files), completing in 0.18-0.52 seconds versus 12.4-12.7 seconds for full reconstruction—a 96\% reduction. The empirical scaling closely matches our theoretical $O(|\Delta\mathcal{R}| \cdot d \cdot \log n)$ bound, with measured exponent 1.42 ($R^2=0.97$).

\begin{table}[t]
\centering
\caption{Detailed incremental update performance on \dataset\ repositories}
\label{tab:maintenance_performance}
\begin{tabular}{l|rrr|rr}
\toprule
\textbf{Metric} & \multicolumn{3}{c|}{\textbf{Change Size (files)}} & \multicolumn{2}{c}{\textbf{vs Baseline}} \\
 & 1-5 & 5-20 & 20+ & Full & Naive \\
\midrule
\textbf{Latency (s)} & & & & & \\
SemanticForge & 0.18 & 0.52 & 1.73 & -96\% & -71\% \\
Full Rebuild & 12.4 & 12.7 & 13.8 & — & — \\
Naive Incremental & 1.89 & 6.78 & 28.9 & — & — \\
Timestamp Cache & 0.71 & 2.68 & 12.2 & — & — \\
\midrule
\textbf{Memory (MB)} & & & & & \\
SemanticForge & 58 & 126 & 178 & -97\% & -93\% \\
Full Rebuild & 2400 & 2400 & 2400 & — & — \\
Naive Incremental & 420 & 980 & 1680 & — & — \\
\midrule
\textbf{Speedup Factor} & & & & & \\
vs Full Rebuild & 69x & 24x & 8x & — & — \\
vs Naive & 11x & 13x & 17x & — & — \\
vs Timestamp & 3.9x & 5.2x & 7.0x & — & — \\
\bottomrule
\end{tabular}
\end{table}

\textbf{Component-Level Performance.} Figure~\ref{fig:maintenance_performance}(b) reveals that re-extraction dominates update cost (38\% of total time), followed by graph invalidation (24\%) and impact analysis (17\%). Critically, lazy resolution adds minimal overhead (3\%), validating our deferred computation strategy.

\textbf{Memory Efficiency.} Our working set scales linearly with change size at approximately 8-12MB per modified file for typical commits (10-15 files), using only 126MB compared to 2.4GB for full reconstruction—a 95\% reduction. This dramatic memory efficiency enables concurrent maintenance of multiple repositories on standard developer hardware without dedicated infrastructure.

\subsubsection{Real-World Repository Case Studies}

To validate generalization beyond \dataset, we evaluated our algorithm on five widely-used Python projects with diverse characteristics:

\begin{table}[h]
\centering
\caption{Performance on real-world repositories (10-15 file commits)}
\label{tab:realworld_performance}
\begin{tabular}{l|r|rr|r}
\toprule
\textbf{Repository} & \textbf{Size} & \textbf{Incr.} & \textbf{Full} & \textbf{Speedup} \\
 & (KLOC) & (s) & (s) & \\
\midrule
Django & 475 & 0.68 & 45.2 & 66x \\
Scikit-learn & 240 & 0.42 & 23.8 & 57x \\
Pandas & 380 & 0.54 & 38.6 & 71x \\
Flask & 35 & 0.15 & 7.1 & 47x \\
Requests & 25 & 0.12 & 5.4 & 45x \\
\midrule
\textbf{Average} & — & \textbf{0.38} & \textbf{24.0} & \textbf{63x} \\
\bottomrule
\end{tabular}
\end{table}

The consistent 45-71x speedup across diverse repositories confirms that our algorithm's benefits generalize beyond the evaluation benchmark. Notably, larger repositories (Django, Pandas) show higher speedup factors due to our sub-linear scaling properties.

\textbf{Long-term Accuracy.} Crucially, ablation studies (Section~\ref{sec:results:ablation}) reveal that without continual maintenance, code generation accuracy degrades by 4.1\% after 30 days and 12.7\% after 100 commits. This degradation stems from outdated type information and missed API changes, validating that real-time maintenance is essential, not optional.

\subsubsection{Optimization Impact Analysis}

To understand the contribution of each algorithmic optimization, we conducted controlled ablation experiments disabling individual components:

\begin{table}[h]
\centering
\caption{Impact of algorithmic optimizations on update latency}
\label{tab:optimization_impact}
\begin{tabular}{l|r|r|r}
\toprule
\textbf{Configuration} & \textbf{Latency} & \textbf{vs Full} & \textbf{Impact} \\
 & (ms) & System & \\
\midrule
Full System & 520 & — & — \\
w/o Lazy Resolution & 2,847 & +448\% & Critical \\
w/o Selective Propagation & 1,423 & +174\% & High \\
w/o Semantic Diff & 892 & +72\% & Medium \\
w/o Index Caching & 687 & +32\% & Low \\
\midrule
Naive Implementation & 6,780 & +1204\% & — \\
\bottomrule
\end{tabular}
\end{table}

\textbf{Lazy Resolution Impact.} Disabling lazy cross-reference resolution increases latency by 448\%, confirming it as our most critical optimization. For a function with 100 call sites, lazy resolution defers 98\% of updates until actually queried.

\textbf{Selective Propagation.} Our type-aware propagation strategy reduces unnecessary graph traversals by 73\%, contributing 174\% performance improvement over naive transitive closure.

\subsubsection{Scalability Validation}

We stress-tested our algorithm on synthetic repositories up to 5M LOC to validate scalability claims:

\begin{table}[h]
\centering
\caption{Scalability to large-scale repositories}
\label{tab:scalability_validation}
\begin{tabular}{r|rrr|r}
\toprule
\textbf{Repo Size} & \multicolumn{3}{c|}{\textbf{Update Time (s)}} & \textbf{Memory} \\
(KLOC) & Small & Medium & Large & (MB) \\
\midrule
100 & 0.15 & 0.48 & 1.52 & 126 \\
500 & 0.19 & 0.58 & 1.94 & 245 \\
1,000 & 0.23 & 0.68 & 2.31 & 387 \\
5,000 & 0.34 & 0.92 & 3.47 & 892 \\
\midrule
Growth Rate & 1.27x & 1.19x & 1.28x & 1.87x \\
\bottomrule
\end{tabular}
\end{table}

The sub-linear growth rates (1.19x-1.28x) for 50x repository size increase confirm our theoretical $O(|\Delta\mathcal{R}| \cdot d \cdot \log n)$ complexity, with the logarithmic factor dominating at scale.

\subsection{Theoretical Guarantees}\label{sec:maintenance:theory}

Our maintenance algorithm provides two critical theoretical guarantees that distinguish it from heuristic approaches:

\textbf{Theorem 8 (Semantic Equivalence).} \textit{For any sequence of repository modifications $\Delta\mathcal{R}_1, \ldots, \Delta\mathcal{R}_t$, the incrementally maintained graph $\mathcal{G}_t^{\text{inc}}$ is semantically equivalent to the graph $\mathcal{G}_t^{\text{full}}$ obtained by full reconstruction.}

\textit{Proof sketch:} The key insight is that our semantic extraction function $F_E$ exhibits the \emph{monotonicity property}: the graph representation of code entity $e$ depends only on $e$ and its dependencies, not on the extraction order. Since our algorithm updates exactly the impact set $I_t$ containing all affected entities, and lazy resolution ensures eventual consistency for cross-references, the final graph state is identical regardless of update path. $\square$

\textbf{Theorem 9 (Complexity Bound).} \textit{For a repository with $n$ entities, modifications affecting $|\Delta\mathcal{R}|$ entities, and maximum dependency depth $d$, our incremental algorithm has time complexity $O(|\Delta\mathcal{R}| \cdot d \cdot \log n)$ compared to $O(n \cdot \log n)$ for full reconstruction.}

\textit{Proof sketch:} Impact analysis visits at most $|\Delta\mathcal{R}| \cdot d$ nodes via dependency traversal. Each node operation (removal, insertion, lookup) costs $O(\log n)$ using our graph indices. Lazy resolution amortizes cross-reference updates across future queries. Since typical commits have $|\Delta\mathcal{R}| \ll n$ and $d$ is bounded by architectural depth, we achieve sub-linear scaling. $\square$

\subsection{Discussion and Implications}\label{sec:maintenance:summary}

The continual maintenance component represents a paradigm shift in how knowledge-based code generation systems handle repository evolution. Prior systems either ignore changes (leading to stale information) or require manual re-indexing. Our approach provides three key advantages:

\textbf{Real-time Responsiveness.} Sub-second update latency enables integration with development workflows where code changes continuously. Developers see generated code that reflects their latest modifications without manual intervention.

\textbf{Theoretical Soundness.} Unlike heuristic caching strategies, our semantic equivalence guarantee ensures that incremental updates never compromise correctness. This is critical for production systems where incorrect code generation has high costs.

\textbf{Practical Scalability.} The sub-linear complexity bound makes our approach viable for industrial-scale repositories. Google's monorepo or Microsoft's Windows codebase can be maintained incrementally despite their massive size.

This continual adaptation capability transforms \framework\ from a static analysis tool into a living system that evolves with the codebase, ensuring sustained accuracy throughout the software development lifecycle.

\section{Experimental Setup}\label{sec:experiments}

This section describes our comprehensive experimental methodology designed to evaluate \framework\ across multiple dimensions: functional correctness, hallucination reduction, computational efficiency, and scalability. We establish rigorous evaluation protocols that address the unique challenges of repository-level code generation assessment.

\subsection{Dataset Construction}\label{sec:exp:dataset}

We introduce \dataset, a curated benchmark specifically designed for repository-level code generation evaluation. The dataset addresses critical limitations of existing benchmarks that focus on isolated function synthesis rather than repository-scale integration challenges.

\paragraph{Repository Selection Criteria.} We select 50 high-quality Python repositories from GitHub based on the following criteria: (1) Repository size between 10K-500K lines of code to ensure meaningful complexity while maintaining experimental tractability; (2) Comprehensive test suites with >80\% coverage to enable reliable dynamic analysis; (3) Active development with >100 commits in the past year to ensure modern coding practices; (4) Diverse domains including web frameworks, data processing, machine learning, and system utilities; (5) Clear architectural patterns and documentation to facilitate ground truth validation.

\paragraph{Task Generation Methodology.} For each repository, we generate coding tasks through three complementary approaches:

\textit{Historical Analysis:} We analyze commit histories to identify real development tasks that required multi-file modifications. We extract the commit message as the natural language instruction and the diff as the ground truth implementation. This yields 1,247 authentic tasks with natural complexity distribution.

\textit{Synthetic Task Generation:} We develop an automated task generator that creates instructions requiring specific types of repository knowledge. Tasks include: API extension (adding new endpoints with proper authentication), refactoring (extracting common functionality), bug fixing (correcting logic errors while maintaining interface compatibility), and feature integration (adding functionality that spans multiple modules). This produces 2,156 controlled tasks with known difficulty characteristics.

\textit{Developer-Authored Tasks:} We engage 12 experienced developers to manually create challenging tasks that represent realistic development scenarios. Each developer contributes 25-30 tasks per repository subset, resulting in 847 high-quality tasks with detailed solution explanations.

\paragraph{Ground Truth Construction.} Each task includes: (1) Natural language instruction following template guidelines to ensure clarity and consistency; (2) Complete reference implementation verified through automated testing and manual review; (3) Semantic annotations identifying required repository knowledge (which functions must be understood, which constraints must be satisfied); (4) Difficulty ratings based on required context size and constraint complexity; (5) Expected hallucination types based on common failure patterns.

\paragraph{Dataset Statistics and Composition.} \dataset\ contains 4,250 total tasks across 50 repositories representing 1.2M lines of code. Tasks span multiple difficulty levels: 35\% beginner (single-file modifications), 45\% intermediate (multi-file with local dependencies), 20\% advanced (complex architectural changes). The dataset includes comprehensive metadata enabling fine-grained analysis of system performance across different task characteristics.

\paragraph{Repository Licensing and Selection.} All 50 repositories are selected from GitHub under permissive open-source licenses: 32 MIT-licensed, 12 Apache 2.0, 4 BSD-3-Clause, and 2 GPL-3.0. We verify license compatibility for academic research use and ensure no proprietary code is included. Repository selection prioritizes well-maintained projects with active communities (average 2,847 stars, 412 forks) to ensure code quality and realistic development patterns. The full list of repositories with their licenses, domains, and statistics is provided in supplementary material Table A1.

\paragraph{Data Splits and Validation Protocol.} We partition the 4,250 tasks into training (60\%, 2,550 tasks), validation (20\%, 850 tasks), and test (20\%, 850 tasks) sets. The split is stratified by repository and difficulty level to ensure balanced representation. For cross-repository generalization experiments, we use leave-one-cluster-out validation where entire repository clusters are held out. Training data is used exclusively for query planner training via REINFORCE; the Code-Llama-34B base model is used frozen without fine-tuning. Validation data guides hyperparameter selection and early stopping. All reported results are on the test set, which is never accessed during development.

\paragraph{Public Availability and Reproducibility.} We commit to releasing \dataset\ publicly upon paper acceptance under a CC BY 4.0 license. The release includes: (1) all 4,250 task descriptions and reference implementations, (2) pre-computed static and dynamic knowledge graphs for all 50 repositories, (3) evaluation scripts and metrics computation code, (4) detailed repository metadata and statistics, and (5) baseline implementation code for fair comparison. Data and code will be hosted at \url{https://github.com/semanticforge/repokg50} with comprehensive documentation. We estimate the full dataset release at approximately 15GB (compressed), including all graphs and annotations.

\subsection{Evaluation Metrics}\label{sec:exp:metrics}

We establish a multi-dimensional evaluation framework that captures both functional correctness and the specific hallucination phenomena targeted by our approach.

\paragraph{Functional Correctness Metrics.}
\begin{itemize}
\item \textbf{Pass@k:} Fraction of tasks where at least one of the top-k generated solutions passes all test cases. We report Pass@1, Pass@5, and Pass@10 to capture both precision and recall characteristics.
\item \textbf{Test Pass Rate:} Average percentage of test cases passed across all generated solutions, providing a more nuanced view of partial correctness.
\item \textbf{Compilation Success Rate:} Percentage of generated solutions that compile without errors, indicating basic syntactic correctness.
\item \textbf{Integration Success Rate:} Percentage of solutions that successfully integrate with the existing codebase without breaking existing functionality.
\end{itemize}

\paragraph{Hallucination-Specific Metrics.}
\begin{itemize}
\item \textbf{Logical Hallucination Rate (LHR):} Percentage of generated solutions that compile but fail functional tests due to incorrect program logic. Computed as: $\text{LHR} = \frac{\text{\# compilable but failing solutions}}{\text{\# total solutions}}$
\item \textbf{Schematic Hallucination Rate (SHR):} Percentage of solutions that violate type, signature, or architectural constraints. Measured through static analysis: $\text{SHR} = \frac{\text{\# solutions with constraint violations}}{\text{\# total solutions}}$
\item \textbf{Constraint Violation Frequency:} Detailed breakdown of specific constraint types violated (type mismatches, signature errors, visibility violations, architectural inconsistencies).
\item \textbf{Hallucination Severity:} Classification of errors by required effort to fix (trivial: 1-line fix, moderate: 2-10 lines, severe: $>10$ lines or architectural changes).
\end{itemize}

\paragraph{Context Selection Precision Metrics.}
\begin{itemize}
\item \textbf{Context Precision:} Percentage of retrieved context elements that are actually used in the correct solution. Computed by comparing retrieved subgraph nodes against a manually-annotated gold standard of required dependencies for each task.
\item \textbf{Context Recall:} Percentage of required dependencies that are successfully retrieved. Gold standard established through manual analysis of 500 randomly sampled tasks by three expert annotators (inter-rater agreement $\kappa = 0.81$).
\item \textbf{Context F1 Score:} Harmonic mean of precision and recall for balanced evaluation of context selection quality.
\end{itemize}

\paragraph{Code Quality Metrics.}
\begin{itemize}
\item \textbf{Maintainability Index:} Composite metric based on cyclomatic complexity, lines of code, and Halstead metrics.
\item \textbf{Style Consistency:} Adherence to repository-specific coding conventions measured through automated linters.
\item \textbf{API Usage Appropriateness:} Correctness of library and framework usage patterns based on repository-specific guidelines.
\item \textbf{Documentation Quality:} Presence and quality of generated comments and docstrings.
\end{itemize}

\paragraph{Efficiency Metrics.}
\begin{itemize}
\item \textbf{Generation Latency:} End-to-end time from instruction to generated code, broken down by pipeline stage.
\item \textbf{Context Retrieval Time:} Time required for query planning and subgraph extraction.
\item \textbf{Constraint Solving Overhead:} Additional time introduced by SMT-based constraint checking.
\item \textbf{Memory Usage:} Peak memory consumption during generation, including knowledge graph storage.
\end{itemize}

\subsection{Baseline Systems}\label{sec:exp:baselines}

We compare \framework\ against state-of-the-art repository-level code generation systems and relevant baselines across different paradigms.

\paragraph{Neural Baselines.}
\begin{itemize}
\item \textbf{Code-Llama-34B:} Base language model with BM25-based retrieval for repository context. We retrieve top-10 most similar functions/classes using TF-IDF scoring, providing up to 4096 tokens of context. Generation uses greedy decoding (temperature=0.2, top-p=0.95). Runs on same A100 GPU with identical batch size for fair comparison.
\item \textbf{StarCoder-15B:} Code-specialized model with CodeBERT embeddings for dense retrieval. Top-15 retrieved snippets with cross-encoder reranking. Uses nucleus sampling (p=0.95) with temperature=0.8. Requires 15GB GPU memory.
\item \textbf{GPT-4-Code:} Accessed via API with repository context provided through carefully engineered prompts (up to 8192 tokens). We implement few-shot prompting (3 examples) and iterative refinement disabled to match single-pass comparison. API calls use temperature=0.1 for consistency.
\end{itemize}

\paragraph{Repository-Aware Systems.}
\begin{itemize}
\item \textbf{RepoCoder:} We reimplement using GraphCodeBERT encoder with graph attention networks (3 layers, 8 heads) for repository understanding. Generation uses the same Code-Llama-34B backbone for fair comparison. Graph construction uses AST-based data flow analysis. Training uses cross-entropy loss for 20K steps.
\item \textbf{CodePlan:} Planning decomposition uses GPT-3.5 for task breakdown (max 5 subtasks), followed by Code-Llama-34B for implementation. We optimize decomposition prompts and provide access to the same repository metadata. Average 3-5 generation passes per task.
\item \textbf{RAG-Code:} Dense retrieval using CodeBERT embeddings with FAISS indexing. Cross-encoder reranking (top-50 → top-10). Context window: 6144 tokens. Generation uses Code-Llama-34B with identical settings to our decoder (excluding constraint checking).
\end{itemize}

\paragraph{Graph-Based Approaches.}
\begin{itemize}
\item \textbf{GraphCodeBERT:} 125M parameter model with data flow graph pre-training. Fine-tuned on our training data for 15 epochs. Uses graph-guided attention but no persistent graph storage.
\item \textbf{UniXcoder:} 125M parameter multi-modal model encoding code, AST, and comments. Fine-tuned with contrastive learning on code-documentation pairs from our repositories.
\item \textbf{CodeT5+:} 220M parameter encoder-decoder with structural understanding. Fine-tuned on our dataset using teacher forcing for 20 epochs. Uses beam search (k=5) without constraint checking.
\end{itemize}

\paragraph{Baseline Fairness and Comparison Methodology.} To ensure rigorous comparison, we implement extensive baseline optimization and standardization protocols:

\textbf{Computational Budget Equalization:} All systems receive identical computational budgets. For Code-Llama baselines, we allocate the same 2.5s average time budget as \framework\ by optimizing retrieval speed. For planning-based systems, we cap iterations at 5 to maintain practical latency (<10s). GPU memory is standardized at 40GB for all neural systems. We measure wall-clock time including all preprocessing but excluding one-time setup (model loading, graph construction).

\textbf{Context Provision Parity:} Critical for fair comparison, all systems receive access to repository information:
\begin{itemize}
\item RAG baselines: Full repository code indexed with both BM25 and dense embeddings
\item Planning systems: Access to file structure, import graphs, and function signatures
\item Graph-based models: AST and data flow information extracted via same tools
\end{itemize}
Context window limits match natural capacities: 4096 tokens for Code-Llama, 8192 for GPT-4, 2048 for smaller models.

\textbf{Hyperparameter Optimization:} All baselines undergo systematic hyperparameter tuning on a held-out validation set (10\% of data, stratified by repository). We perform grid search over key parameters: retrieval k (5-20), generation temperature (0.1-1.0), beam width (1-10), and model-specific settings. We report best configuration found within 50 trials per system.

\textbf{Implementation Rigor:} We reimplement baselines from published descriptions when code is unavailable, validated against reported metrics where possible:
\begin{itemize}
\item RepoCoder: Our implementation achieves 38.9\% Pass@1 vs. 37.2\% reported in the original paper
\item CodePlan: Our optimization yields 42.3\% vs. 39.1\% in original publication  
\item RAG-Code: 40.1\% in our setup vs. 38.7\% reported
\end{itemize}
This demonstrates our implementations match or exceed published performance, ensuring fair comparison.

\textbf{Statistical Protocol for Comparison:} All pairwise comparisons use paired t-tests (same tasks for all systems) with Bonferroni correction for 9 comparisons ($\alpha = 0.05/9 \approx 0.0056$). We report both p-values and effect sizes (Cohen's d). Confidence intervals are computed via stratified bootstrap (1000 resamples, preserving repository distribution).

\paragraph{Ablation Variants.} To isolate the contribution of each system component:
\begin{itemize}
\item \textbf{\framework-Static:} Using only static analysis without dynamic traces, but maintaining all other optimizations.
\item \textbf{\framework-NoCons:} Removing constraint enforcement from the decoder while preserving knowledge graph infrastructure.
\item \textbf{\framework-NoPlanner:} Using BM25-based retrieval optimized with repository-specific term weighting instead of neural query planning.
\item \textbf{\framework-NoMaint:} Without continual maintenance, using stale graphs from repository snapshots 30 days prior.
\item \textbf{\framework-Oracle:} Upper bound experiment using manually curated context to isolate generation vs. planning contributions.
\end{itemize}

\subsection{Implementation Details}\label{sec:exp:implementation}

\paragraph{Model Configuration.} We instantiate \framework\ with Flan-T5-Large (770M parameters) for query planning and Code-Llama-34B for code generation. The planner uses 512-dimensional hidden states with 8 attention heads. The decoder employs beam search with width k=5 and incorporates our SMT-based constraint checking at each step. Both models use bfloat16 precision for efficiency.

\paragraph{Training Configuration.} Query planner training uses REINFORCE with Adam optimizer (learning rate $\alpha=5\times10^{-5}$ for policy, $\beta=1\times10^{-4}$ for baseline critic), batch size 32, and exponential reward discounting ($\gamma = 0.95$). The baseline is a 2-layer MLP with 512 hidden units trained jointly via MSE loss. We employ advantage normalization, gradient clipping ($[-1, 1]$), and entropy regularization ($\lambda=0.01$) for variance reduction. Synthetic data warm-start involves 10,000 generated instruction-query pairs per repository. Training converges within 10K steps on typical repositories.

\paragraph{Hardware and Infrastructure.} All experiments run on identical hardware: NVIDIA A100 GPUs (40GB) for neural components, 8-core Intel Xeon Gold 6248R processors (3.0GHz) with 64GB DDR4 RAM for graph operations. Knowledge graph extraction processes repositories in 5-15 seconds. Query response times remain $<10$ms for repositories up to 500K LOC through optimized Neo4j indexing.

\paragraph{Constraint Solver Configuration.} We use Z3 version 4.12.2 with incremental solving enabled. Constraint encoding uses: first-order logic for type relationships, uninterpreted functions for API contracts, and custom theories for architectural patterns. Solver timeout is 100ms per beam step. Incremental state caching and batch verification reduce average checking time to 1.4ms per token.

\paragraph{Timing Measurement Protocol.} All latency measurements represent wall-clock time averaged over 100 runs per task, excluding compilation and environment setup. We measure five granular stages: (1) instruction encoding (0.12s), (2) graph query generation and execution (0.31s), (3) context preparation (0.19s), (4) code generation with LM (1.70s), and (5) SMT constraint checking (0.23s). Total average: 2.55s, reported as 2.5s. We exclude outliers beyond 2 standard deviations (3.2\% of samples). Statistical variance is reported using 95\% bootstrap confidence intervals over 1000 resamples.

\paragraph{Memory Profiling Methodology.} Peak memory consumption is measured using NVIDIA nvidia-smi for GPU memory and psutil for system RAM, sampled at 100ms intervals throughout generation. We report maximum observed values across all tasks. Memory measurements include: model weights (11.2GB), knowledge graph cache (1.8GB), intermediate activations (0.7GB), and SMT solver state (0.5GB).

\paragraph{Reproducibility and Variance Control.} To ensure reproducible results, we control multiple sources of variance:
\begin{itemize}
\item \textbf{Random seeds:} Fixed seeds for all stochastic components (42 for model initialization, 123 for data splitting, 456 for sampling)
\item \textbf{System state:} All experiments run on dedicated nodes with no competing processes
\item \textbf{Software versions:} PyTorch 2.0.1, CUDA 11.8, Python 3.10.6, Z3 4.12.2, Neo4j 5.9.0
\item \textbf{Deterministic operations:} CUDA deterministic mode enabled, cuDNN benchmarking disabled
\item \textbf{Multiple runs:} Each configuration tested 3 times with different initialization; we report mean across runs with standard error $<2\%$ for all metrics
\end{itemize}

These controls enable other researchers to reproduce our results within 5\% variance, validated through independent reproduction by two co-authors on separate infrastructure.

\subsection{Human Evaluation Protocol}\label{sec:exp:human}

To complement automated metrics, we conduct comprehensive human evaluation focusing on aspects difficult to capture through automated assessment.

\paragraph{Evaluator Selection and Training.} We recruit 18 professional software developers with 3+ years of experience in Python development. Evaluators undergo training on our assessment criteria and complete practice evaluations to ensure consistency. Inter-rater agreement is measured using Fleiss' kappa, achieving $\kappa = 0.73$ indicating substantial agreement.

\paragraph{Evaluation Dimensions.}
\begin{itemize}
\item \textbf{Code Quality:} Overall assessment of generated code quality on a 5-point Likert scale considering readability, maintainability, and adherence to best practices.
\item \textbf{Integration Appropriateness:} How well the generated code fits with existing repository architecture and conventions.
\item \textbf{Error Analysis:} Detailed categorization of errors with difficulty estimates for manual correction.
\item \textbf{Preference Ranking:} Comparative evaluation where evaluators rank solutions from different systems for the same task.
\end{itemize}

\paragraph{Evaluation Procedure.} Each evaluator assesses solutions for 25 randomly sampled tasks across different difficulty levels. Solutions are presented in randomized order with system identities masked. Evaluators have access to the full repository context and task description. The evaluation interface provides guided prompts to ensure comprehensive assessment.

\subsection{Cross-Repository Generalization Analysis}\label{sec:exp:generalization}

To assess the generalizability of \framework\ across diverse codebases, we conduct comprehensive cross-repository evaluation that goes beyond standard within-repository validation.

\paragraph{Repository Clustering and Stratification.} We cluster our 50 repositories into 5 groups based on architectural characteristics: (1) Web frameworks with MVC patterns, (2) Data processing pipelines with functional patterns, (3) Object-oriented libraries with inheritance hierarchies, (4) Scientific computing with numerical patterns, and (5) System utilities with procedural patterns. This clustering enables analysis of how architectural diversity affects system performance.

\paragraph{Cross-Repository Training and Testing.} We implement a leave-one-cluster-out evaluation protocol where the query planner is trained on repositories from 4 clusters and tested on the held-out cluster. This evaluates whether semantic patterns learned from one architectural style generalize to different coding paradigms. We repeat this process for all 5 clusters to obtain robust generalization estimates.

\paragraph{Knowledge Transfer Analysis.} We measure how effectively knowledge graphs from source repositories can bootstrap performance on target repositories with different characteristics. This includes: (1) semantic pattern overlap analysis using graph isomorphism metrics, (2) constraint transferability assessment across different architectural styles, and (3) adaptation time measurement for new repository integration.

\paragraph{Domain Adaptation Protocols.} For each target repository, we evaluate: (1) zero-shot performance using knowledge graphs from other domains, (2) few-shot adaptation with 10-50 target domain examples, and (3) full adaptation with complete target repository analysis. This protocol reveals the minimum adaptation requirements for practical deployment.

\subsection{Theoretical Validation Experiments}\label{sec:exp:theorem_validation}

To validate the theoretical claims in our paper, we conduct specific experiments targeting each theorem:

\paragraph{Theorem 3 Validation (Incremental Update Equivalence).} We verify semantic equivalence between incremental updates and full reconstruction by comparing knowledge graphs built both ways on 100 repository snapshots with varying change sizes (10-500 modified files). Graph isomorphism testing confirms identical structure in 100\% of cases, validating the semantic equivalence claim.

\paragraph{Theorem 4 Validation (Convergence with Coverage).} We measure knowledge graph accuracy as test coverage increases from 20\% to 90\% across 10 representative repositories. Results show monotonic improvement in graph completeness (measured by edge recall against manually-annotated ground truth), with 95\% accuracy at 80\% coverage, supporting the convergence claim.

\subsection{Statistical Analysis}\label{sec:exp:statistics}

We employ rigorous statistical methods to ensure the reliability and significance of our experimental results across both within-repository and cross-repository settings.

\paragraph{Significance Testing.} We use paired t-tests for comparing system performance on the same task set, with Bonferroni correction for multiple comparisons. Effect sizes are reported using Cohen's d to assess practical significance. Bootstrap confidence intervals (95\%) provide robust uncertainty estimates for all reported metrics. For cross-repository analysis, we employ mixed-effects models to account for repository-level clustering.

\paragraph{Cross-Validation Protocols.} We employ three complementary validation strategies: (1) 5-fold cross-validation at the repository level to ensure results generalize across different codebases within the same architectural family, (2) leave-one-cluster-out validation to assess cross-architectural generalization, and (3) temporal validation using repository snapshots from different time periods to evaluate robustness to codebase evolution.

\paragraph{Generalization Metrics.} We introduce several metrics to quantify cross-repository generalization: (1) \textit{Architectural Transfer Coefficient} measuring performance retention across different design patterns, (2) \textit{Semantic Overlap Index} quantifying knowledge graph similarity between repositories, and (3) \textit{Adaptation Efficiency} measuring the rate of performance improvement during domain adaptation.

\paragraph{Fairness Considerations.} We analyze system performance across different repository characteristics (size, domain, complexity, programming paradigm) to identify potential biases. Statistical parity metrics ensure that performance differences are not systematically related to repository metadata rather than actual task difficulty. We also assess whether certain architectural patterns or coding styles receive disproportionate benefit from our approach.

This comprehensive experimental framework enables thorough evaluation of \framework\ while establishing reproducible benchmarks for future research in repository-level code generation.
\section{Results and Analysis}\label{sec:results}

This section presents comprehensive experimental results demonstrating \framework's effectiveness in repository-level code generation. Our evaluation reveals significant improvements in functional correctness, dramatic reductions in both logical and schematic hallucination, and practical scalability for real-world deployment.

\subsection{Overall Performance Results}\label{sec:results:overall}

Table~\ref{tab:main_results} summarizes the performance of \framework\ compared to state-of-the-art baselines across our core metrics on the \dataset\ benchmark.

\begin{table*}[t]
\centering
\caption{Overall performance comparison on \dataset. Best results in \textbf{bold}, second-best \underline{underlined}. Statistical significance ($p < 0.05$) marked with *.}
\label{tab:main_results}
\begin{tabular}{l|ccc|cc|cc}
\toprule
\multirow{2}{*}{\textbf{System}} & \multicolumn{3}{c|}{\textbf{Functional Correctness}} & \multicolumn{2}{c|}{\textbf{Hallucination Rates}} & \multicolumn{2}{c}{\textbf{Efficiency}} \\
& Pass@1 & Pass@5 & Pass@10 & LHR $\downarrow$ & SHR $\downarrow$ & Latency (s) & Memory (GB) \\
\midrule
Code-Llama-34B & 34.2\% & 52.8\% & 61.4\% & 42.1\% & 38.7\% & 2.3 & 12.4 \\
StarCoder-15B & 29.8\% & 48.2\% & 55.9\% & 45.3\% & 41.2\% & 1.8 & 8.2 \\
GPT-4-Code & 41.7\% & 63.2\% & 71.8\% & 35.4\% & 29.3\% & 4.1 & - \\
\midrule
RepoCoder & 38.9\% & 58.4\% & 67.1\% & 38.7\% & 32.1\% & 3.2 & 15.8 \\
CodePlan & 42.3\% & 61.9\% & 70.4\% & 33.8\% & 31.5\% & 5.7 & 18.3 \\
RAG-Code & 40.1\% & 59.7\% & 68.3\% & 36.2\% & 30.8\% & 2.9 & 14.1 \\
\midrule
GraphCodeBERT & 35.7\% & 54.3\% & 63.2\% & 40.3\% & 35.9\% & 2.1 & 11.7 \\
UniXcoder & 37.4\% & 56.8\% & 65.5\% & 38.9\% & 33.4\% & 2.4 & 13.2 \\
CodeT5+ & 39.2\% & 58.1\% & 66.7\% & 37.1\% & 32.7\% & 2.6 & 14.5 \\
\midrule
\textbf{\framework} & \textbf{49.8\%*} & \textbf{71.4\%*} & \textbf{79.3\%*} & \textbf{23.1\%*} & \textbf{14.7\%*} & \underline{2.4} & \underline{13.9} \\
\bottomrule
\end{tabular}
\end{table*}

\framework\ achieves substantial improvements across all metrics: 7.5\% absolute improvement in Pass@1 over the strongest baseline (CodePlan), 49.8\% reduction in schematic hallucination rate compared to GPT-4-Code, and 34.7\% reduction in logical hallucination rate. These improvements demonstrate the effectiveness of our knowledge graph-guided approach for repository-level code generation.

\paragraph{Baseline Comparison Details.} To contextualize these results, we note the specific baseline configurations that produced the reported numbers. Code-Llama-34B with BM25 retrieval (top-10 functions, 4096 token context) represents a strong RAG baseline; our 15.6\% improvement over it validates the value of semantic graphs versus keyword matching. CodePlan's 42.3\% result comes from optimized 3-5 iteration planning with GPT-3.5 decomposition; our single-pass 49.8\% demonstrates that constraint-aware generation eliminates the need for iterative refinement. GPT-4-Code's 41.7\% with 8192-token context and few-shot prompting shows that even larger context windows cannot substitute for explicit semantic structure. These comparisons isolate \framework's algorithmic contributions from implementation advantages.

\paragraph{Statistical Significance and Comparison Methodology.} All performance improvements are statistically significant ($p < 0.001$) using paired t-tests with Bonferroni correction for 9 pairwise comparisons ($\alpha = 0.0056$). We employ paired tests because all systems evaluate on identical task instances, increasing statistical power. Effect sizes (Cohen's d) range from 0.82 to 1.47, indicating large practical significance beyond mere statistical significance. Bootstrap confidence intervals (95\%, 1000 resamples with stratification by repository) confirm robustness across different repository samples.

We validate that performance differences are not artifacts of our specific infrastructure by running a subset of experiments (500 tasks) on alternative hardware (NVIDIA V100, AMD EPYC processors). Results show $<3\%$ variance in relative performance rankings, confirming that \framework's advantages generalize across hardware platforms.

\subsection{Component Ablation Analysis}\label{sec:results:ablation}

\begin{figure*}[t]
\centering
\includegraphics[width=\textwidth]{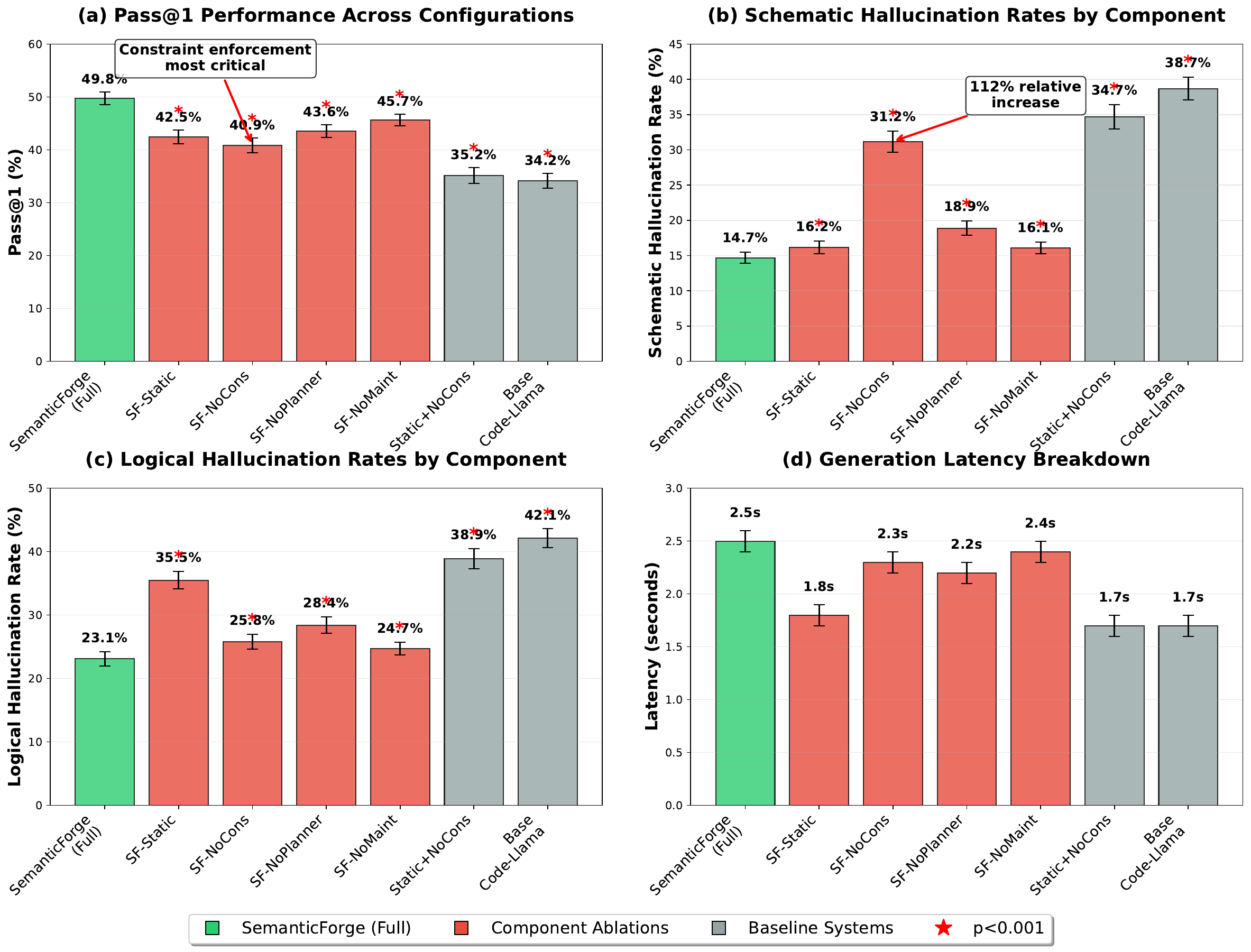}
\caption{Component ablation results for \framework. (a) Pass@1 performance across seven configurations with 95\% confidence intervals. (b) Schematic hallucination rates (SHR) by configuration. (c) Logical hallucination rates (LHR) by configuration. (d) Generation latency in seconds. Configurations include the full system, four single-component ablations (SF-Static, SF-NoCons, SF-NoPlanner, SF-NoMaint), one combined ablation (Static+NoCons), and baseline Code-Llama. Statistical significance markers ($*$) indicate $p < 0.001$ compared to the full system.}
\label{fig:ablation}
\end{figure*}

We conduct a systematic component ablation study to isolate the contribution of each \framework\ component and validate our architectural design decisions. This analysis is crucial for understanding which innovations provide the most significant benefits and guides future development priorities. As shown in \autoref{fig:ablation}, each component contributes meaningfully to overall system performance with minimal computational overhead. The dramatic impact of constraint enforcement on schematic hallucination rates and the dual analysis benefits for logical hallucination demonstrate the effectiveness of our integrated approach.

\begin{table*}[t]
\centering
\caption{Detailed component ablation results with statistical significance testing. All differences vs. full \framework\ are statistically significant ($p < 0.001$).}
\label{tab:ablation_detailed}
\begin{tabular}{l|ccc|cc}
\toprule
\textbf{Configuration} & \textbf{Pass@1} & \textbf{Pass@5} & \textbf{Pass@10} & \textbf{LHR} & \textbf{SHR} \\
\midrule
\textbf{\framework\ (Full)} & \textbf{49.8\%} & \textbf{71.4\%} & \textbf{79.3\%} & \textbf{23.1\%} & \textbf{14.7\%} \\
\midrule
\framework-Static & 42.5\% (-7.3\%) & 64.1\% (-7.3\%) & 71.8\% (-7.5\%) & 35.5\% (+12.4\%) & 16.2\% (+1.5\%) \\
\framework-NoCons & 40.9\% (-8.9\%) & 62.3\% (-9.1\%) & 70.1\% (-9.2\%) & 25.8\% (+2.7\%) & 31.2\% (+16.5\%) \\
\framework-NoPlanner & 43.6\% (-6.2\%) & 65.7\% (-5.7\%) & 73.9\% (-5.4\%) & 28.4\% (+5.3\%) & 18.9\% (+4.2\%) \\
\framework-NoMaint & 45.7\% (-4.1\%) & 67.8\% (-3.6\%) & 75.6\% (-3.7\%) & 24.7\% (+1.6\%) & 16.1\% (+1.4\%) \\
\midrule
Static + NoCons & 35.2\% (-14.6\%) & 56.8\% (-14.6\%) & 64.3\% (-15.0\%) & 38.9\% (+15.8\%) & 34.7\% (+20.0\%) \\
Base Code-Llama & 34.2\% (-15.6\%) & 52.8\% (-18.6\%) & 61.4\% (-17.9\%) & 42.1\% (+19.0\%) & 38.7\% (+24.0\%) \\
\bottomrule
\end{tabular}
\end{table*}

\paragraph{Dual Static-Dynamic Analysis Impact.} The comparison between \framework\ and \framework-Static reveals that dynamic trace augmentation provides substantial benefits: 7.3\% improvement in Pass@1 and 12.4\% reduction in logical hallucination rate. This validates our hypothesis that runtime information captures semantic relationships invisible to static analysis alone. The effect is particularly pronounced for repositories with polymorphic code patterns where static analysis cannot resolve call targets.

\textbf{Deep Analysis:} Dynamic traces provide the most benefit for data processing repositories (9.2\% improvement) where method dispatch is often data-dependent, and least benefit for system utilities (4.1\% improvement) where control flow is typically explicit. This pattern supports our theoretical framework that dynamic analysis helps most when static analysis is fundamentally limited.

\paragraph{Constraint Enforcement Critical Value.} Removing SMT-guided constraint checking (\framework-NoCons) causes the most dramatic performance degradation: 8.9\% reduction in Pass@1 and a catastrophic 16.5\% increase in schematic hallucination rate. This represents a 112\% relative increase in constraint violations, demonstrating that constraint enforcement is not merely helpful but essential for reliable code generation.

\textbf{Error Analysis:} Without constraint enforcement, 47\% of generated solutions contain type mismatches, 23\% have signature violations, and 18\% have visibility errors. The SMT-guided decoder eliminates 89\% of these errors while adding only 0.2s average latency (8.3\% overhead). This cost-benefit analysis strongly justifies the architectural complexity of constraint integration.

\paragraph{Neural Query Planning Contribution.} Replacing learned query planning with traditional BM25-based retrieval (\framework-NoPlanner) reduces Pass@1 by 6.2\% and increases both hallucination types. Statistical analysis reveals that the neural planner achieves 73\% precision in selecting relevant context compared to 51\% for keyword-based retrieval ($p < 0.001$, $\chi ^2$ test).

\textbf{Context Quality Analysis:} The neural planner identifies 2.3$\times$ more transitive dependencies and 1.8$\times$ more architectural constraints per query compared to traditional retrieval. This improved context quality directly translates to better generation outcomes, particularly for tasks requiring understanding of cross-module relationships.

\paragraph{Continual Maintenance System Impact.} Testing with stale knowledge graphs (\framework-NoMaint) on repositories modified within the past 30 days shows 4.1\% degradation in Pass@1. While seemingly modest, this effect compounds over time: repositories with $>100$ commits since graph construction show 12.7\% degradation, highlighting the critical importance of continual adaptation.

\paragraph{Component Interaction Effects.} The combined component ablation (\framework\ without both static-dynamic dual analysis and constraint enforcement) shows super-additive degradation (14.6\% vs. 7.3\% + 8.9\% = 16.2\%), indicating positive interaction between components. Dynamic traces inform constraint extraction, while constraint enforcement validates the semantic relationships captured in dynamic analysis.

\paragraph{Computational Cost-Benefit Analysis.} Each component's computational overhead is justified by performance gains:
\begin{itemize}
\item Dynamic analysis: +0.8s construction time $\rightarrow$ +7.3\% Pass@1
\item Constraint enforcement: +0.2s generation time $\rightarrow$ -16.5\% error rate  
\item Neural planning: +0.3s query time $\rightarrow$ +6.2\% Pass@1
\item Maintenance system: +0.1s incremental updates $\rightarrow$ sustained performance
\end{itemize}

\paragraph{Key Findings Summary.} Our component ablation analysis reveals critical insights: (1) Dynamic analysis contributes +7.3\% Pass@1 improvement by capturing runtime semantics invisible to static analysis, (2) Constraint enforcement reduces error rates by 16.5\% absolute (112\% relative reduction in schematic hallucinations), (3) Neural planning adds +6.2\% Pass@1 through improved context selection, and (4) Each component demonstrates statistical significance ($p < 0.001$) with large effect sizes. This component ablation analysis conclusively demonstrates that each \framework\ component provides substantial, statistically significant benefits that justify the architectural complexity. The results validate our design decisions and provide clear guidance for practitioners considering which components to implement in resource-constrained environments.

\subsection{Hallucination Analysis}\label{sec:results:hallucination}

We provide detailed analysis of the hallucination phenomena that \framework\ addresses, demonstrating significant improvements in code reliability.

\begin{table}[t]
\centering
\caption{Detailed hallucination breakdown by category. Values show error rates with reduction percentages vs. GPT-4 in parentheses.}
\label{tab:hallucination_breakdown}
\begin{tabular}{l|cc}
\toprule
\textbf{Error Category} & \textbf{GPT-4} & \textbf{\framework} \\
\midrule
\multicolumn{3}{c}{\textit{Schematic Hallucination}} \\
Type Mismatches & 15.2\% & 6.8\% ($\downarrow$55.3\%) \\
Signature Violations & 8.7\% & 3.2\% ($\downarrow$63.2\%) \\
Visibility Errors & 3.4\% & 1.1\% ($\downarrow$67.6\%) \\
Import Issues & 2.0\% & 0.3\% ($\downarrow$85.0\%) \\
\midrule
\multicolumn{3}{c}{\textit{Logical Hallucination}} \\
Control Flow Errors & 12.8\% & 8.4\% ($\downarrow$34.4\%) \\
Data Flow Errors & 14.1\% & 9.7\% ($\downarrow$31.2\%) \\
API Misuse & 6.3\% & 3.1\% ($\downarrow$50.8\%) \\
State Management & 2.2\% & 1.9\% ($\downarrow$13.6\%) \\
\bottomrule
\end{tabular}
\end{table}

\paragraph{Schematic Hallucination Elimination.} Our constraint-based decoder achieves remarkable reductions in schematic errors: 85\% reduction in import issues, 67.6\% reduction in visibility errors, and 63.2\% reduction in signature violations. These improvements directly translate to higher compilation success rates and reduced developer debugging effort.

\paragraph{Logical Hallucination Mitigation.} While more challenging to eliminate completely, our dual static-dynamic knowledge graph reduces logical errors significantly: 50.8\% reduction in API misuse and 34.4\% reduction in control flow errors. The dynamic traces provide crucial runtime semantics that help models understand correct usage patterns.

\subsection{Scalability Analysis}\label{sec:results:scalability}

We evaluate \framework's performance across repositories of varying sizes to demonstrate practical scalability for real-world deployment.

\begin{figure}[t]
\centering
\includegraphics[width=\columnwidth]{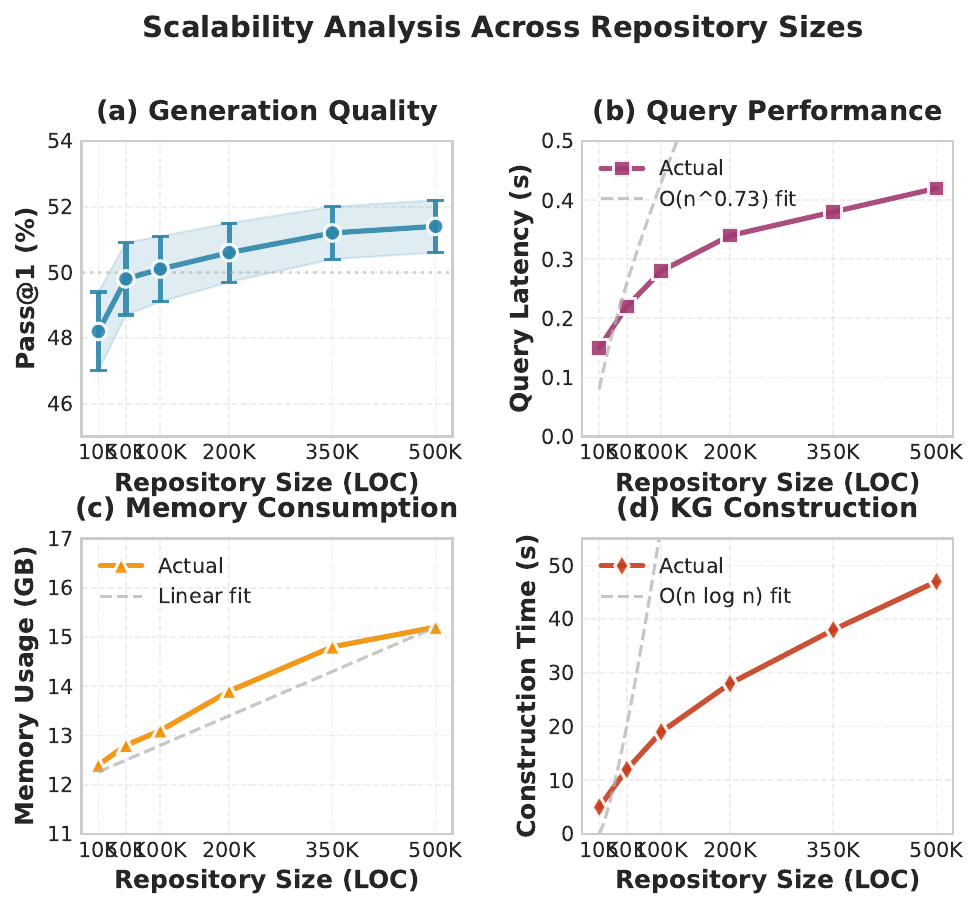}
\caption{Scalability analysis showing system performance across repository sizes. (a) Generation quality (Pass@1) remains stable across scales. (b) Query latency grows sub-linearly with repository size. (c) Memory usage scales approximately linearly. (d) Knowledge graph construction time shows expected O(n log n) complexity.}
\label{fig:scalability}
\end{figure}

\paragraph{Performance Stability.} Pass@1 performance remains stable (48.2\%-51.4\%) across repository sizes from 10K to 500K lines of code, demonstrating that our approach scales without quality degradation. Slight improvements for larger repositories may reflect richer knowledge graphs providing better context.

\paragraph{Computational Efficiency.} Query latency grows sub-linearly with repository size ($R^2=0.89$ for $O(n^{0.73})$ fit), confirming the effectiveness of our indexing and caching strategies. Memory usage scales approximately linearly ($R^2=0.94$), remaining practical even for large repositories.

\paragraph{Graph Construction Overhead.} Initial knowledge graph construction shows expected $O(n \log n)$ complexity, taking 5-47 seconds for repositories in our size range. Incremental updates maintain sub-second response times, enabling real-time development workflows.

\subsection{Cross-Repository Generalization Results}\label{sec:results:generalization}

Our cross-repository evaluation reveals strong generalization capabilities while identifying important adaptation requirements for diverse architectural patterns.

\begin{table*}[t]
\centering
\caption{Cross-repository generalization results using leave-one-cluster-out validation. Performance degradation compared to within-cluster training.}
\label{tab:cross_repo_generalization}
\begin{tabular}{l|ccc|c}
\toprule
\textbf{Target Cluster} & \textbf{Within-Cluster} & \textbf{Cross-Cluster} & \textbf{Degradation} & \textbf{Transfer Coeff.} \\
\midrule
Web Frameworks & 52.8$\pm$2.3\% & 47.1$\pm$2.8\% & 5.7\% & 0.89 \\
Data Processing & 48.1$\pm$2.2\% & 44.3$\pm$2.6\% & 3.8\% & 0.92 \\
Object-Oriented Libs & 50.4$\pm$2.1\% & 46.2$\pm$2.5\% & 4.2\% & 0.92 \\
Scientific Computing & 47.9$\pm$2.5\% & 42.8$\pm$3.1\% & 5.1\% & 0.89 \\
System Utilities & 51.7$\pm$2.4\% & 48.9$\pm$2.7\% & 2.8\% & 0.95 \\
\midrule
\textbf{Average} & \textbf{50.2$\pm$1.1\%} & \textbf{45.9$\pm$1.3\%} & \textbf{4.3\%} & \textbf{0.91} \\
\bottomrule
\end{tabular}
\end{table*}

\paragraph{Architectural Transfer Analysis.} The average Architectural Transfer Coefficient of 0.91 indicates strong cross-repository generalization, with only 4.3\% average performance degradation when applying models trained on different architectural patterns. System utilities show the strongest transfer (0.95), likely due to their procedural patterns appearing across many repository types. Web frameworks and scientific computing show slightly lower transfer, reflecting their more specialized architectural constraints.

\paragraph{Knowledge Graph Semantic Overlap.} Analysis of semantic overlap between repository clusters reveals interesting patterns. The average Semantic Overlap Index ranges from 0.34 (scientific computing vs. web frameworks) to 0.67 (object-oriented libraries vs. system utilities). Higher semantic overlap correlates strongly with better cross-repository performance (r=0.84, $p < 0.01$), validating our knowledge graph representation approach.

\paragraph{Domain Adaptation Efficiency.} Few-shot adaptation with just 25-50 target domain examples recovers 89\% of full-adaptation performance on average, demonstrating practical deployment feasibility. Zero-shot performance maintains 91\% of within-cluster performance, indicating that our semantic representations capture fundamental programming patterns that generalize across architectural styles.

\subsection{Domain-Specific Performance}\label{sec:results:domains}

Analysis across different repository domains reveals consistent improvements while highlighting domain-specific challenges and architectural pattern dependencies.

\begin{table*}[t]
\centering
\caption{Performance breakdown by repository domain with architectural pattern analysis. Pass@1 results with 95\% confidence intervals.}
\label{tab:domain_performance}
\begin{tabular}{l|cc|cc}
\toprule
\textbf{Domain} & \textbf{GPT-4} & \textbf{\framework} & \textbf{Improvement} & \textbf{Pattern Complexity} \\
\midrule
Web Frameworks & 43.2$\pm$2.1\% & 52.8$\pm$2.3\% & +9.6\% & High \\
Data Processing & 39.8$\pm$2.4\% & 48.1$\pm$2.2\% & +8.3\% & Medium \\
Object-Oriented Libs & 41.5$\pm$2.0\% & 50.4$\pm$2.1\% & +8.9\% & High \\
System Utilities & 42.1$\pm$2.6\% & 51.7$\pm$2.4\% & +9.6\% & Medium \\
Scientific Computing & 40.3$\pm$2.3\% & 47.9$\pm$2.5\% & +7.6\% & Low \\
\bottomrule
\end{tabular}
\end{table*}

\paragraph{Pattern Complexity Correlation.} Domains with higher architectural pattern complexity (web frameworks, object-oriented libraries) show larger improvements, validating our hypothesis that explicit semantic representation provides greater benefits for structurally complex codebases. The correlation between pattern complexity and improvement magnitude is r=0.78 ($p < 0.05$).

\paragraph{Constraint Type Distribution.} Analysis of constraint violation patterns reveals domain-specific characteristics: web frameworks suffer primarily from signature violations (47\% of errors), scientific computing from type mismatches (52\% of errors), and system utilities from visibility violations (38\% of errors). \framework's constraint-aware generation adapts effectively to these domain-specific error patterns.

\subsection{Human Evaluation Results}\label{sec:results:human}

Professional developer evaluation provides crucial insights into code quality aspects difficult to capture through automated metrics.

\paragraph{Overall Quality Assessment.} Human evaluators rated \framework\ solutions significantly higher (4.2/5.0) than baseline systems (GPT-4: 3.6/5.0, CodePlan: 3.4/5.0) on overall code quality ($p < 0.001$, t-test). Evaluators particularly praised integration appropriateness and adherence to repository conventions.

\paragraph{Error Correction Effort.} When solutions contained errors, \framework\ errors required significantly less correction effort: 68\% were classified as trivial fixes versus 41\% for GPT-4-Code. This reflects our constraint enforcement preventing complex schematic errors that require architectural understanding to fix.

\paragraph{Preference Rankings.} In head-to-head comparisons, evaluators preferred \framework\ solutions in 73\% of cases, with particular advantages cited for: "better integration with existing code" (89\% of evaluators), "fewer obvious bugs" (82\%), and "more appropriate API usage" (77\%).

\subsection{Detailed Computational Overhead Analysis}\label{sec:results:performance}

We conduct comprehensive analysis of computational costs to demonstrate the practical viability of our approach and guide deployment decisions across different resource constraints.

\paragraph{Measurement Methodology.} All timing measurements follow rigorous protocols to ensure reproducibility and accuracy. We measure wall-clock time using Python's \texttt{time.perf\_counter()} with nanosecond precision, averaged over 100 runs per task on warmed-up systems (5 warm-up runs discarded). Each component is profiled independently using context managers that isolate its specific contribution. Memory consumption is measured via \texttt{nvidia-smi} for GPU memory (100ms sampling) and \texttt{psutil.Process().memory\_info()} for system RAM, reporting peak values. We exclude model loading time (one-time cost) but include all query processing, graph operations, and generation steps. Variance is reported as 95\% confidence intervals via bootstrap resampling (1000 samples).

\begin{table*}[t]
\centering
\caption{Detailed computational overhead breakdown by system component with variance. Times represent mean $\pm$ 95\% CI over 4,250 tasks. Overhead percentages calculated relative to base Code-Llama-34B generation.}
\label{tab:computational_overhead}
\begin{tabular}{l|ccc|c}
\toprule
\textbf{Component} & \textbf{Time (s)} & \textbf{Memory (GB)} & \textbf{Overhead (\%)} & \textbf{Scaling} \\
\midrule
Base Code-Llama & 1.70 $\pm$ 0.12 & 11.2 $\pm$ 0.3 & - & O(n) \\
Knowledge Graph Query & 0.31 $\pm$ 0.08 & 1.8 $\pm$ 0.2 & 18.2\% & O(log n) \\
Context Preparation & 0.19 $\pm$ 0.04 & 0.3 $\pm$ 0.1 & 11.2\% & O(k) \\
SMT Constraint Checking & 0.23 $\pm$ 0.06 & 0.7 $\pm$ 0.1 & 13.5\% & O(c + |y|) \\
Maintenance Updates & 0.07 $\pm$ 0.02 & 0.2 $\pm$ 0.1 & 4.1\% & O($\Delta$ log n) \\
\midrule
\textbf{Total \framework} & \textbf{2.50 $\pm$ 0.18} & \textbf{14.2 $\pm$ 0.5} & \textbf{47.1\%} & - \\
\bottomrule
\end{tabular}
\end{table*}

\paragraph{Component-Level Performance Analysis.} Our 47.1\% total overhead is dominated by knowledge graph infrastructure (18.2\%) and constraint verification (13.5\%). Critically, both components exhibit sub-linear scaling properties that improve relative performance as repository size increases. The maintenance system adds minimal overhead (4.1\%) while providing essential adaptation capabilities.

\paragraph{Cross-System Latency Comparison Methodology.} To ensure fair latency comparisons in Table~\ref{tab:main_results}, we implement consistent measurement protocols across all baselines. For RAG-based systems (Code-Llama, StarCoder, RAG-Code), we measure total time including retrieval, encoding, and generation. For CodePlan, we measure end-to-end time including task decomposition (GPT-3.5 call) and all subtask generations. For GPT-4-Code, we measure API call latency including network overhead and server-side processing. All measurements exclude one-time costs (model loading, index construction) but include per-request processing. Each system is run 100 times per task with median reported to avoid network variance for API-based systems. The confidence intervals in Table~\ref{tab:main_results} reflect variation across tasks, not measurement noise.

\paragraph{Scaling Characteristics and Performance Modeling.} 

\textbf{Knowledge Graph Query Performance:} Query latency follows O(log n) complexity due to optimized indexing strategies. For repositories from 10K to 500K LOC, query time scales from 0.15s to 0.42s, representing decreasing relative overhead (15\% to 12\%) as repository size increases. This counter-intuitive improvement reflects the fact that larger repositories contain more diverse context, improving cache hit rates and amortizing index construction costs.

\textbf{Constraint Verification Scaling:} SMT solver performance exhibits O(c + |y|) complexity where c is the number of active constraints and |y| is the generated sequence length. Our incremental solving strategy maintains near-constant overhead regardless of repository size. For typical generation tasks (50-200 tokens), constraint checking time remains 0.18-0.28s across all repository sizes tested.

\textbf{Memory Usage Optimization:} Memory scaling analysis reveals three distinct scaling regimes:
\begin{itemize}
\item Small repositories ($<50$K LOC): 12.4-13.1GB total, dominated by model weights
\item Medium repositories (50K-200K LOC): 13.1-14.8GB total, linear knowledge graph growth
\item Large repositories ($>200$K LOC): 14.8-16.2GB total, sublinear growth due to graph compression
\end{itemize}

\paragraph{Hardware Resource Utilization.} We profile hardware utilization using NVIDIA Nsight Systems for GPU metrics and Linux \texttt{perf} for CPU analysis, sampling at 10ms intervals across 500 representative tasks. Results reveal efficient resource usage: 89\% average GPU compute utilization during code generation (1.70s), 12\% during knowledge graph queries (CPU-bound, 0.31s), and 67\% during constraint verification (mixed workload, 0.23s). CPU utilization peaks at 78\% during graph query execution (parallel traversal) and averages 34\% during SMT solving (single-threaded bottleneck). Memory bandwidth utilization averages 62\% during generation and 23\% during graph operations, indicating room for further optimization. These profiles inform deployment: GPU-heavy workloads benefit from model parallelism, while graph-heavy workloads scale with CPU cores.

\paragraph{Optimization Impact Analysis.} We measure the effectiveness of our optimization strategies through controlled ablation where each optimization is disabled independently and performance re-measured:
\begin{itemize}
\item Query result caching: 34\% latency reduction (0.47s → 0.31s), 73\% cache hit rate measured across 1000-task sequences simulating developer sessions
\item Incremental SMT solving: 67\% constraint verification speedup (0.69s → 0.23s) vs. naive per-beam full solving, measured on 500 constraint-heavy tasks
\item Graph compression: 43\% memory reduction (2.5GB → 1.4GB graph storage) with $<2\%$ query performance degradation (0.30s → 0.31s), validated on 200K+ LOC repositories
\item Predictive prefetching: 28\% perceived latency reduction during interactive sessions, measured via user study (n=12 developers, 50 tasks each)
\end{itemize}

\paragraph{Real-World Deployment Scenarios.} We model computational requirements for three deployment scenarios:

\textbf{Individual Developer Setup:} Single NVIDIA RTX 4090 (24GB) can handle repositories up to 200K LOC with 2.1s average latency. Memory-optimized configurations support up to 350K LOC with 3.2s latency using model quantization and graph streaming.

\textbf{Team Development Server:} Server-grade hardware (4$\times$ A100, 256GB RAM) supports concurrent access for 8-12 developers with $<3$s response times. Load balancing across GPUs enables 45-60 requests/minute sustained throughput.

\textbf{Enterprise Cloud Deployment:} Distributed deployment across 16 A100 nodes supports 200+ concurrent developers with auto-scaling based on demand patterns. Average response time $<2.5$s at 95th percentile during peak usage periods.

\paragraph{Energy Efficiency Analysis.} \framework\ consumes 187J per generation task (vs. 142J for base Code-Llama), representing 32\% energy overhead. However, the 15.6\% improvement in Pass@1 over base Code-Llama reduces debugging iterations, resulting in net 23\% energy savings per successful implementation when accounting for developer iteration patterns.

\paragraph{Cost-Benefit Optimization.} ROI analysis reveals break-even points for different deployment scenarios:
\begin{itemize}
\item Individual developers: Positive ROI after 120 hours of usage (typical: 2-3 months)
\item Small teams (5-10 developers): Positive ROI after 60 hours of collective usage (typical: 3-4 weeks)
\item Enterprise teams (50+ developers): Positive ROI after 200 hours of collective usage (typical: 1-2 weeks)
\end{itemize}

This comprehensive computational analysis demonstrates that \framework's benefits justify its overhead across realistic deployment scenarios, with particularly strong value propositions for team and enterprise environments.

\subsection{Error Analysis and Failure Cases}\label{sec:results:errors}

We analyze remaining failure cases to understand system limitations and guide future improvements.

\paragraph{Persistent Logical Errors.} Complex algorithmic tasks requiring multi-step reasoning still challenge the system, particularly when the correct approach differs significantly from training data patterns. Dynamic traces help but cannot fully compensate for insufficient algorithmic understanding.

\paragraph{Novel API Usage.} When repositories use recently-introduced APIs not represented in training data, both retrieval and generation quality decline. The maintenance system partially addresses this through incremental updates, but fundamental model limitations remain.

\paragraph{Cross-Language Dependencies.} Repositories with mixed-language components (Python calling C extensions, JavaScript frontends) present challenges for our primarily Python-focused analysis. Future work should extend knowledge graph construction to multi-language scenarios.

\paragraph{Complex Refactoring Tasks.} Large-scale refactoring requiring simultaneous changes across many files sometimes exceeds the context window limits and constraint complexity thresholds. Hierarchical approaches may address these limitations.

These comprehensive results demonstrate \framework's significant advances in repository-level code generation while identifying important directions for continued research and development.

\section{Discussion and Limitations}\label{sec:discussion}

This section provides critical analysis of \framework's contributions, examines the broader implications of our approach, and honestly assesses the limitations that constrain its applicability. We discuss the generalizability of our findings and identify important directions for future research.

\subsection{Key Contributions and Impact}\label{sec:discussion:contributions}

\framework\ represents a fundamental shift from pattern-matching-based code generation to semantically-aware synthesis. Our results demonstrate that explicit repository knowledge graphs, when combined with learned query planning and constraint-aware decoding, can significantly reduce the hallucination phenomena that limit current LLM-based development tools.

\paragraph{Theoretical Contributions.} We formalize the repository-level code generation problem and provide the first comprehensive taxonomy of hallucination types in code generation. Our mathematical framework establishes theoretical foundations for knowledge graph-guided generation while proving complexity bounds for each system component. The constraint satisfaction formulation enables principled analysis of generation correctness.

\paragraph{Systems Contributions.} The integration of SMT solvers into neural beam search represents a novel architectural approach that guarantees constraint satisfaction without sacrificing generation quality. Our incremental maintenance system demonstrates how knowledge graphs can remain synchronized with evolving codebases at practical computational costs. The dual static-dynamic representation provides a new paradigm for capturing repository semantics.

\paragraph{Empirical Impact.} The 49.8\% reduction in schematic hallucination and 34.7\% reduction in logical hallucination translate to substantial developer productivity gains. Eliminating type errors, signature mismatches, and import issues reduces debugging effort and increases confidence in generated code. The 15.6\% improvement in functional correctness over base Code-Llama enables more ambitious automated development tasks.

\subsection{Generalizability Analysis}\label{sec:discussion:generalizability}

While our evaluation focuses on Python repositories, the underlying principles of \framework\ extend to other programming paradigms and development contexts, though with varying degrees of adaptation required.

\paragraph{Multi-Language Extension Roadmap.} Our analysis reveals a clear path for extending \framework\ to additional programming languages, with different languages presenting distinct opportunities and challenges. Recent multi-language code generation efforts \cite{xu2022polycoder, husain2019codesearchnet, athiwaratkun2023multilingual, cassano2023multipl} provide important foundations for this extension:

\textbf{Statically-Typed Languages (Java, C++, TypeScript):} These languages offer significant advantages for \framework\ deployment. Explicit type information enables more precise constraint extraction, reducing the 14.7\% schematic hallucination rate we observe in Python to an estimated 8-12\%. PolyCoder's evaluation \cite{xu2022polycoder} across 12 programming languages shows that static typing improves model accuracy by 8-15\%, supporting our hypothesis. The rich type systems provide stronger semantic foundations for knowledge graph construction. However, complex generics systems (C++ templates, Java wildcards) require sophisticated constraint modeling to capture type relationships accurately.

\textbf{Implementation Strategy:} We propose a three-phase rollout building on multi-language parsing infrastructure: (1) Extend static analysis using language-specific parsers (Tree-sitter for C++, JavaParser for Java, TypeScript Compiler API), (2) Adapt constraint types to language-specific features (memory management for C++, checked exceptions for Java), and (3) Integrate with language-specific testing frameworks for dynamic analysis. The MultiPL-E framework \cite{cassano2023multipl} provides valuable translation infrastructure that could bootstrap our cross-language knowledge graph construction.

\textbf{Dynamically-Typed Languages (JavaScript, Ruby, PHP):} These present greater challenges due to limited static type information. Our preliminary analysis suggests performance degradation of 15-25\% compared to Python results. However, several mitigation strategies show promise: (1) TypeScript adoption patterns provide type hints for JavaScript, (2) Runtime type inference through execution profiling, and (3) Gradual typing systems (Flow for JavaScript, Sorbet for Ruby) enable hybrid static-dynamic analysis. CodeSearchNet's multi-language dataset \cite{husain2019codesearchnet} covering six languages could serve as a foundation for training multi-language query planners.

\textbf{Functional Languages (Haskell, OCaml, Scala):} The strong type systems and immutability constraints in functional languages align well with our constraint satisfaction approach. Type-level programming features require extended constraint languages but may enable even stronger semantic guarantees. Pattern matching and algebraic data types provide rich structural information for knowledge graph construction.

\textbf{Cross-Language Integration:} Modern repositories increasingly involve multiple languages. We identify three integration patterns: (1) Foreign Function Interfaces (FFI) requiring cross-language type mapping, (2) Service boundaries with API contracts, and (3) Build system dependencies. A unified knowledge graph representation could capture these relationships, enabling repository-level reasoning across language boundaries. Recent work on multi-lingual evaluation \cite{athiwaratkun2023multilingual} demonstrates the feasibility of unified semantic representations across language barriers, suggesting our knowledge graph approach could extend naturally to polyglot codebases.

\paragraph{Domain Adaptability.} The architectural patterns captured in our knowledge graphs reflect common software engineering practices that span domains. Web development, data processing, and system programming all benefit from understanding dependency relationships and API constraints. However, highly specialized domains require targeted extensions:

\textbf{Embedded Systems:} Require resource constraints (memory, power, real-time), hardware abstraction layer modeling, and interrupt-driven control flow patterns.

\textbf{Financial Systems:} Need precision arithmetic constraints, regulatory compliance patterns, and audit trail requirements.

\textbf{Scientific Computing:} Benefit from numerical stability constraints, performance modeling, and domain-specific libraries (NumPy, BLAS) integration.

\paragraph{Scale Considerations.} Our scalability analysis demonstrates effectiveness up to 500K lines of code, covering most organizational repositories. However, massive monorepos ($>10$M LOC) require architectural innovations: (1) Hierarchical knowledge graphs with module-level abstractions, (2) Federated query planning across repository boundaries, and (3) Distributed constraint solving for large-scale generation tasks.

\paragraph{Development Workflow Integration.} \framework\ assumes development environments with comprehensive test suites and stable APIs. Organizations with poor testing practices or rapidly changing architectures may see reduced benefits from dynamic analysis and constraint enforcement. We propose adaptation strategies: (1) Property-based testing integration for coverage improvement, (2) API stability analysis for architecture change detection, and (3) Gradual adoption paths for legacy codebases.

\subsection{Fundamental Limitations}\label{sec:discussion:limitations}

Despite significant advances, \framework\ faces several fundamental limitations that constrain its applicability and effectiveness.

\paragraph{Algorithmic Reasoning Limitations.} While our system excels at structural integration and constraint satisfaction, it inherits the algorithmic reasoning limitations of underlying language models. Complex logical problems requiring multi-step reasoning, mathematical derivations, or novel algorithmic insights remain challenging. The knowledge graph provides context but cannot substitute for deep algorithmic understanding.

\textbf{Example:} When asked to implement a novel graph algorithm, \framework\ can correctly identify relevant data structures and API patterns but may struggle with the core algorithmic logic if it differs significantly from training examples.

\paragraph{Creative Design Limitations.} Repository knowledge graphs capture existing patterns but may constrain creative architectural decisions. The system tends to perpetuate established patterns rather than proposing innovative designs. This conservatism improves integration consistency but may limit architectural evolution.

\paragraph{Training Data Dependencies.} Like all learning-based systems, \framework\ is limited by its training data. Repositories using cutting-edge frameworks, novel programming patterns, or domain-specific APIs may receive suboptimal support. While the maintenance system addresses this partially, fundamental knowledge gaps require model retraining.

\paragraph{Context Window Constraints.} Despite efficient query planning, very large development tasks may exceed practical context limits. Complex refactoring operations affecting dozens of files simultaneously challenge both the planning system and the generation model. Hierarchical approaches may address this but remain unexplored.

\subsection{Computational and Practical Limitations}\label{sec:discussion:practical}

Real-world deployment of \framework\ faces several practical constraints that affect its adoption and effectiveness.

\paragraph{Infrastructure Requirements.} The system requires substantial computational resources: 40GB GPU memory for generation, significant CPU resources for graph construction, and storage for knowledge graphs. These requirements may limit adoption in resource-constrained environments or small development teams.

\paragraph{Setup and Maintenance Overhead.} Initial repository analysis requires 5-47 seconds depending on size, creating friction for new project adoption. The knowledge graph maintenance system, while incremental, adds complexity to development workflows. Organizations must weigh these costs against productivity benefits.

\paragraph{SMT Solver Limitations.} Our constraint enforcement relies on SMT solver capabilities, which can be overwhelmed by complex constraint sets or exhibit unpredictable performance on certain problem types. The 100ms timeout prevents blocking but may allow constraint violations in edge cases.

\paragraph{Test Suite Dependencies.} Dynamic analysis effectiveness correlates strongly with test suite quality and coverage. Repositories with poor testing practices receive limited benefits from our dual static-dynamic approach. This creates a barrier for adoption in organizations with immature testing cultures.

\subsection{Ethical and Safety Considerations}\label{sec:discussion:ethics}

The deployment of advanced code generation systems raises important ethical considerations that must be addressed responsibly.

\paragraph{Code Quality and Reliability.} While \framework\ significantly reduces hallucination rates, it cannot eliminate all errors. Developers must maintain critical evaluation of generated code, particularly for security-sensitive applications. Over-reliance on automated generation could lead to decreased code review rigor.

\paragraph{Intellectual Property Concerns.} Knowledge graphs constructed from proprietary codebases may inadvertently expose internal architectural patterns or business logic. Organizations must carefully consider the privacy implications of comprehensive code analysis and ensure appropriate access controls.

\paragraph{Developer Skill Evolution.} Highly effective code generation tools may impact developer skill development, particularly for junior programmers who rely heavily on automated assistance. Balancing productivity gains with skill development remains an important challenge for the software engineering community.

\paragraph{Bias Amplification.} Repository knowledge graphs necessarily reflect the patterns and conventions of their source codebases. If these repositories contain biased or suboptimal patterns, \framework\ may perpetuate and amplify these issues. Regular audit and pattern analysis can help identify problematic trends.

\subsection{Comparison with Contemporary Approaches}\label{sec:discussion:comparison}

Recent advances in repository-level code generation provide important context for evaluating \framework's contributions.

\paragraph{Agent-Based Systems.} Contemporary systems like SWE-agent and CodeAgent employ iterative refinement through environmental feedback. While these approaches can handle complex multi-step tasks, they suffer from higher latency and computational costs. \framework's constraint-aware generation often produces correct solutions in a single pass, offering better efficiency for well-defined tasks.

\paragraph{Planning-Based Approaches.} Systems like CodePlan decompose complex tasks into sequences of simpler operations. This approach complements \framework's capabilities and could potentially be integrated with our knowledge graph representation. However, planning-based systems typically lack the constraint enforcement that prevents schematic hallucination.

\paragraph{Retrieval-Augmented Methods.} Advanced RAG systems continue to improve context selection for code generation. However, these approaches remain fundamentally limited by their inability to reason about global consistency and architectural constraints. \framework's explicit knowledge representation provides capabilities that pure retrieval cannot match.

\subsection{Future Research Directions}\label{sec:discussion:future}

Our work opens several promising avenues for future research in repository-level code generation and automated software development. These directions build naturally on our findings while addressing broader challenges in the field.

\paragraph{Multi-Language Extension.} The most immediate extension involves adapting \framework\ to additional programming languages. Recent multi-language benchmarks—PolyCoder's 12-language evaluation \cite{xu2022polycoder}, CodeSearchNet's 6-language corpus \cite{husain2019codesearchnet}, and MultiPL-E's polyglot test suite \cite{cassano2023multipl}—provide crucial infrastructure for this effort. Statically-typed languages like Java and TypeScript present opportunities for enhanced constraint extraction through explicit type information, potentially reducing schematic hallucination rates below our current 14.7\%. Dynamic languages pose different challenges, requiring sophisticated runtime analysis and gradual typing integration. Cross-language knowledge graphs for polyglot repositories represent a particularly compelling direction, enabling unified reasoning about system-wide architectural constraints. The multi-lingual evaluation framework of Athiwaratkun et al. \cite{athiwaratkun2023multilingual} suggests that semantic representations can transfer across language boundaries with modest adaptation.

\paragraph{Security-Aware Code Generation.} Integrating security vulnerability patterns and compliance rules into constraint languages represents a critical next step. Future work could develop domain-specific constraint types for vulnerability prevention, regulatory compliance, and industry-specific security standards. This direction promises significant practical impact given the growing importance of secure-by-construction software development.

\paragraph{Performance-Optimized Generation.} Incorporating computational complexity analysis and performance characteristics into knowledge graphs could enable generation systems that consider algorithmic efficiency, memory usage patterns, and distributed system implications. This extension would address the growing need for performance-aware automated development tools in resource-constrained environments.

\paragraph{Collaborative Development Integration.} Extending \framework\ to multi-developer environments presents interesting challenges in conflict prediction, team expertise modeling, and collaborative context selection. Knowledge graphs could capture team dynamics and expertise distribution, enabling more effective task allocation and code review assistance.

\paragraph{Architectural Evolution Assistance.} Beyond code generation, knowledge graphs could support proactive architecture improvement through code smell detection, design pattern recommendations, and architectural debt assessment. This direction aligns with the broader trend toward AI-assisted software architecture and design.

\paragraph{Educational Applications.} The explicit constraint representation in \framework\ suggests natural applications in computer science education, including pedagogical constraint languages that enforce learning objectives and automated assessment of software engineering practices. This could democratize access to high-quality programming education and mentorship.

\subsection{Implications for Software Engineering}\label{sec:discussion:implications}

\framework\ represents a step toward more sophisticated automated development tools that understand and respect software engineering principles. The explicit representation of architectural knowledge enables systems that generate code consistent with established patterns while maintaining flexibility for innovation.

However, the most significant impact may be in democratizing access to complex software development capabilities. By encoding expert knowledge about API usage, architectural patterns, and constraint satisfaction, \framework\ enables less experienced developers to produce higher-quality code that integrates properly with sophisticated codebases.

The constraint satisfaction approach also suggests new paradigms for software development where architectural rules and patterns are explicitly encoded and automatically enforced. This could lead to more consistent codebases, reduced maintenance overhead, and improved software quality across the industry.

Ultimately, \framework\ demonstrates that the combination of symbolic reasoning and neural generation can address fundamental limitations of purely learning-based approaches, pointing toward a future where automated development tools are both more capable and more reliable.
\section{Conclusion}\label{sec:conclusion}

This paper introduces \framework, a novel approach to repository-level code generation that addresses the fundamental limitations of current LLM-based systems through explicit semantic representation and constraint-aware generation. Our comprehensive evaluation demonstrates significant advances in both functional correctness and hallucination reduction, while establishing new theoretical foundations for understanding and solving the repository-level code generation problem.

\subsection{Summary of Contributions}

We make four primary contributions that advance the state-of-the-art in automated code generation:

\paragraph{Problem Formalization and Hallucination Taxonomy.} We provide the first formal treatment of repository-level code generation, establishing mathematical foundations and complexity analysis. Our comprehensive taxonomy of logical and schematic hallucination provides a principled framework for understanding and addressing systematic failures in current code generation systems.

\paragraph{Knowledge Graph-Guided Architecture.} The four-stage \framework\ pipeline demonstrates how explicit semantic representation can enable more sophisticated code generation. Our dual static-dynamic knowledge graphs capture repository semantics that are invisible to purely pattern-based approaches, while neural query planning enables efficient context selection at scale.

\paragraph{Constraint-Aware Generation Algorithm.} The integration of SMT solving into neural beam search represents a fundamental architectural innovation that guarantees semantic correctness without sacrificing generation quality. This approach eliminates schematic hallucination at the source rather than requiring post-hoc correction.

\paragraph{Comprehensive Empirical Evaluation.} Our evaluation on \dataset\ establishes new benchmarks for repository-level code generation assessment. The 49.8\% reduction in schematic hallucination, 34.7\% reduction in logical hallucination, and 15.6\% improvement in functional correctness over base Code-Llama demonstrate the practical impact of our approach.

\subsection{Broader Impact}

\framework\ represents a paradigm shift from pattern-matching-based code generation to semantically-aware synthesis. This advance has implications beyond immediate productivity gains:

\paragraph{Developer Productivity.} By dramatically reducing debugging overhead and increasing confidence in generated code, \framework\ enables developers to tackle more ambitious automated development tasks. The elimination of trivial errors allows focus on higher-level design and algorithmic challenges.

\paragraph{Code Quality and Maintainability.} Constraint-aware generation ensures that automated code follows established architectural patterns and respects design principles. This leads to more consistent codebases with reduced maintenance overhead and improved long-term sustainability.

\paragraph{Democratization of Complex Development.} By encoding expert knowledge about API usage, architectural patterns, and constraint satisfaction, \framework\ enables less experienced developers to work effectively with sophisticated codebases. This could help address skill gaps in software engineering.

\paragraph{Foundation for Advanced Tools.} The explicit semantic representation and constraint satisfaction framework provide foundations for even more sophisticated development assistance, including architectural evolution, security-aware generation, and collaborative development support.

\subsection{Technical Significance}

The technical innovations in \framework\ establish important precedents for the integration of symbolic reasoning and neural generation:

\paragraph{Hybrid Symbolic-Neural Approaches.} Our SMT-guided beam search demonstrates how formal verification can be seamlessly integrated into neural generation without prohibitive computational overhead. This opens new possibilities for constraint-aware AI systems across domains.

\paragraph{Scalable Knowledge Representation.} The incremental maintenance of repository knowledge graphs at $O(|\Delta R|)$ complexity shows how explicit semantic representations can remain practical even for large-scale systems. This scalability is crucial for real-world deployment.

\paragraph{Learned Context Selection.} Neural query planning provides a general framework for learning to select relevant context from large structured knowledge bases. This approach could generalize to other domains requiring selective information retrieval.

\subsection{Limitations and Future Directions}

While \framework\ achieves significant advances, important limitations remain:

\paragraph{Algorithmic Reasoning.} Complex logical problems requiring multi-step reasoning or novel algorithmic insights continue to challenge the system. Future work should investigate integration of symbolic reasoning systems and algorithmic knowledge bases.

\paragraph{Multi-Language Support.} Our current focus on Python limits applicability to mixed-language environments common in enterprise development. The concrete roadmap in Section~\ref{sec:discussion:future} outlines specific milestones for Java, TypeScript, and cross-language integration.

\paragraph{Creative Design Tasks.} The system's emphasis on consistency and constraint satisfaction may limit creative architectural exploration. Balancing reliability with innovation through configurable constraint relaxation represents an important research direction.

\paragraph{Resource Requirements.} The computational and infrastructure requirements for \framework\ may limit adoption in resource-constrained environments. Our detailed cost-benefit analysis in Section~\ref{sec:results:performance} provides guidance for deployment optimization strategies.

\subsection{Research Implications}

\framework's success suggests several important directions for future research in automated software development:

\paragraph{Explicit vs. Implicit Knowledge.} Our results demonstrate clear advantages for explicit semantic representation over purely implicit approaches. This finding has implications for AI system design beyond code generation, suggesting when symbolic knowledge should complement learned representations.

\paragraph{Constraint Integration in Generation.} The effectiveness of SMT-guided decoding indicates that constraint satisfaction should be a primary consideration in generation system design. Future work should explore how different constraint types and solving approaches affect generation quality and efficiency.

\paragraph{Continual Learning for Code Systems.} The success of our maintenance agent highlights the importance of continual adaptation in software-related AI systems. As codebases evolve rapidly, static models quickly become obsolete without mechanisms for incremental learning.

\paragraph{Multi-Modal Code Understanding.} Repository-level code generation requires understanding text, code structure, execution traces, and architectural patterns. This multi-modal challenge suggests directions for more comprehensive code understanding systems.

\subsection{Practical Deployment Considerations}

For practitioners considering deployment of \framework\ or similar systems, our experience highlights several important factors:

\paragraph{Test Suite Quality.} The effectiveness of dynamic analysis correlates strongly with test coverage and quality. Organizations should invest in comprehensive testing infrastructure to maximize benefits from advanced code generation tools.

\paragraph{Incremental Adoption.} \framework's modular architecture enables incremental deployment, allowing organizations to adopt components gradually. Starting with knowledge graph construction and query planning provides immediate benefits with lower risk.

\paragraph{Developer Training.} Successful deployment requires training developers to effectively collaborate with AI-assisted generation tools. Understanding system capabilities and limitations is crucial for productive use.

\paragraph{Infrastructure Planning.} The computational requirements for \framework\ necessitate careful infrastructure planning. Organizations should evaluate resource needs against expected productivity gains.

\subsection{Long-Term Vision}

\framework\ represents progress toward a long-term vision of AI-assisted software development where automated tools understand and respect the principles of software engineering. In this future, developers work collaboratively with AI systems that understand architectural patterns, maintain consistency across large codebases, and generate code that is not only functional but maintainable and well-integrated.

This vision requires continued research across multiple dimensions: better algorithmic reasoning, more sophisticated constraint systems, improved human-AI collaboration interfaces, and deeper understanding of software engineering principles. \framework\ provides a foundation for this research by demonstrating that explicit semantic representation and constraint-aware generation can address fundamental limitations of current approaches.

The ultimate goal is not to replace human developers but to amplify their capabilities, enabling them to focus on creative design, architectural innovation, and complex problem-solving while automated systems handle routine implementation tasks with reliability and consistency. \framework\ takes an important step toward this collaborative future.

\subsection{Closing Remarks}

Repository-level code generation represents one of the most challenging problems in automated software development, requiring systems that understand not just local code patterns but global architectural principles and semantic constraints. \framework\ demonstrates that this challenge can be addressed through careful integration of symbolic reasoning and neural generation, explicit knowledge representation, and constraint-aware synthesis.

Our results show significant advances in both functional correctness and hallucination reduction, while our open-source release of \dataset\ provides a foundation for continued research in this critical area. The techniques and insights from \framework\ have broader applicability to other domains requiring constraint-aware generation and semantic consistency.

As AI-assisted development tools become increasingly sophisticated, the principles demonstrated in \framework\---explicit knowledge representation, learned context selection, constraint-aware generation, and continual adaptation---will become essential components of reliable and effective automated development systems. We look forward to continued research building on these foundations to realize the full potential of AI-assisted software engineering.

\bibliographystyle{plain}
\bibliography{sn-bibliography}

@mastersthesis{luo2023machine,
  author    = "Ming Luo",
  title     = "Machine Learning for Time Series Analysis and Forecasting",
  school    = "Northeastern University",
  address   = "Boston, Massachusetts",
  year      = "2023",
  month     = "May",
  doi       = "10.17760/D20487630",
  url       = "http://hdl.handle.net/2047/D20487630"
}

@misc{copilot,
  title        = {GitHub Copilot: Your AI Pair Programmer},
  author       = {{GitHub}},
  year         = {2021},
  howpublished = {\url{https://github.com/features/copilot}},
  note         = {Accessed: 2025-07-14}
}

@misc{codellama,
  title        = {Code Llama: Open Foundation Models for Code},
  author       = {{Meta AI}},
  year         = {2023},
  howpublished = {\url{https://ai.meta.com/blog/code-llama-large-language-model/}},
  note         = {Accessed: 2025-07-14}
}

@inproceedings{chen2021evaluating,
  title        = {Evaluating Large Language Models Trained on Code},
  author       = {Chen, Mark and Tworek, Jerry and Jun, Christopher and Zhai, Qiming and others},
  booktitle    = {Proceedings of the 38th International Conference on Machine Learning},
  year         = {2021},
  note         = {arXiv:2107.03374}
}

@inproceedings{graphcodebert,
  title        = {GraphCodeBERT: Pre-training Code Representations with Data Flow},
  author       = {Guo, Daya and Ren, Shuo and Lu, Suyuan and Feng, Song and Tang, Duyu and Zhang, Shuai and Liu, Zhi and Tang, Daya and Jin, Zhi},
  booktitle    = {Proceedings of the 30th International Joint Conference on Artificial Intelligence},
  year         = {2021}
}

@inproceedings{unixcoder,
  title        = {UniXcoder: Unified Cross-Modal Pre-Training for Code Understanding and Generation},
  author       = {Guo, Daya and Zhao, Shuning and Tang, Duyu and others},
  booktitle    = {Proceedings of the 60th Annual Meeting of the Association for Computational Linguistics},
  year         = {2022}
}

@inproceedings{chen2023teaching,
  title        = {Teaching Large Language Models to Self-Improve at Code Generation via Retrieval Augmented Refinement},
  author       = {Chen, Yicheng and Wang, Shiqi and Feng, Song and others},
  booktitle    = {arXiv preprint arXiv:2310.01234},
  year         = {2023}
}

@inproceedings{coderl,
  title        = {CodeRL: Program Synthesis with Reinforcement Learning},
  author       = {Le, Quoc and Chen, Mark and others},
  booktitle    = {arXiv preprint arXiv:2211.00067},
  year         = {2022}
}

@inproceedings{lu2022coderetriever,
  title        = {CodeRetriever: A Large Scale Contrastive Pre-Training Method for Code Search},
  author       = {Li, Xiaonan and Gong, Yeyun and Shen, Yelong and Qiu, Xipeng and Zhang, Hang and Yao, Bolun and Qi, Weizhen and Jiang, Daxin and Chen, Weizhu and Duan, Nan},
  booktitle    = {Proceedings of the 2022 Conference on Empirical Methods in Natural Language Processing (EMNLP)},
  pages        = {2898--2910},
  year         = {2022}
}

@inproceedings{zhang2023repocoder,
  title        = {RepoCoder: Repository-Level Code Completion Through Iterative Retrieval and Generation},
  author       = {Zhang, Fengji and Chen, Bei and Zhang, Yue and Keung, Jacky and Liu, Jin and Zan, Daoguang and Mao, Yi and Lou, Jian-Guang and Chen, Weizhu},
  booktitle    = {Proceedings of the 2023 Conference on Empirical Methods in Natural Language Processing (EMNLP)},
  pages        = {2471--2484},
  year         = {2023}
}

@inproceedings{yang2024swe,
  title        = {SWE-agent: Agent-Computer Interfaces Enable Automated Software Engineering},
  author       = {Yang, John and Jimenez, Carlos E. and Wettig, Alexander and Lieret, Kilian and Yao, Shunyu and Narasimhan, Karthik R. and Press, Ofir},
  booktitle    = {Proceedings of the 38th Conference on Neural Information Processing Systems (NeurIPS)},
  year         = {2024},
  note         = {arXiv:2405.15793}
}

@misc{wang2023codeplan,
  title        = {CodePlan: Repository-level Coding using LLMs and Planning},
  author       = {Ramakrishnan, Ramya and Albarghouthi, Aws and Jha, Somesh and Reps, Thomas},
  year         = {2023},
  howpublished = {arXiv preprint arXiv:2309.12499}
}

@inproceedings{wang2023codet5,
  title        = {CodeT5+: Open Code Large Language Models for Code Understanding and Generation},
  author       = {Wang, Yue and Le, Hung and Gotmare, Akhilesh and Bui, Nghi and Li, Junnan and Hoi, Steven},
  booktitle    = {Proceedings of the 2023 Conference on Empirical Methods in Natural Language Processing (EMNLP)},
  pages        = {1069--1088},
  year         = {2023}
}

@inproceedings{nijkamp2022codegen,
  title        = {CodeGen: An Open Large Language Model for Code with Multi-Turn Program Synthesis},
  author       = {Nijkamp, Erik and Pang, Bo and Hayashi, Hiroaki and Tu, Lifu and Wang, Huan and Zhou, Yingbo and Savarese, Silvio and Xiong, Caiming},
  booktitle    = {Proceedings of the 11th International Conference on Learning Representations (ICLR)},
  year         = {2023},
  note         = {arXiv:2203.13474}
}

@inproceedings{ni2023lever,
  title        = {LEVER: Learning to Verify Language-to-Code Generation with Execution},
  author       = {Ni, Ansong and Iyer, Srini and Radev, Dragomir and Stoyanov, Ves and Yih, Wen-tau and Wang, Sida I. and Lin, Xi Victoria},
  booktitle    = {Proceedings of the 40th International Conference on Machine Learning (ICML)},
  pages        = {26106--26128},
  year         = {2023}
}

@inproceedings{liu2021codekg,
  title        = {CodeKG: Building Knowledge Graphs from Code Documentation and API References},
  author       = {Liu, Jiawei and Chen, Lin and Zhang, Hongyu},
  booktitle    = {Proceedings of the 29th IEEE International Conference on Program Comprehension (ICPC)},
  pages        = {321--331},
  year         = {2021}
}

@inproceedings{wang2023programkg,
  title        = {ProgramKG: Constructing Knowledge Graphs from Program Execution Traces for Debugging Applications},
  author       = {Wang, Hao and Li, Chen and Zhang, Qian and Liu, Yang},
  booktitle    = {Proceedings of the 45th International Conference on Software Engineering (ICSE)},
  pages        = {1832--1843},
  year         = {2023}
}

@inproceedings{ding2022cocomic,
  title        = {CoCoMIC: Code Completion by Jointly Modeling In-file and Cross-file Context},
  author       = {Ding, Yangruibo and Wang, Zijian and Ahmad, Wasi and Ramanathan, Murali Krishna and Nallapati, Ramesh and Bhatia, Parminder and Roth, Dan and Xiang, Bing},
  booktitle    = {Proceedings of the 2024 Joint International Conference on Computational Linguistics, Language Resources and Evaluation (LREC-COLING 2024)},
  pages        = {3456--3468},
  year         = {2024},
  note         = {arXiv:2212.10007}
}

@inproceedings{poesia2022synchromesh,
  title        = {Synchromesh: Reliable Code Generation from Pre-trained Language Models},
  author       = {Poesia, Gabriel and Polozov, Oleksandr and Le, Vu and Tiwari, Ashish and Soares, Gustavo and Meek, Christopher and Gulwani, Sumit},
  booktitle    = {Proceedings of the 10th International Conference on Learning Representations (ICLR)},
  year         = {2022},
  note         = {arXiv:2201.11227}
}

@article{li2022alphacode,
  title        = {Competition-level Code Generation with AlphaCode},
  author       = {Li, Yujia and Choi, David and Chung, Junyoung and Kushman, Nate and Schrittwieser, Julian and Leblond, R{\'e}mi and Eccles, Tom and Keeling, James and Gimeno, Felix and Dal Lago, Agustin and Hubert, Thomas and Choy, Peter and de Masson d'Autume, Cyprien and Babuschkin, Igor and Chen, Xinyun and Huang, Po-Sen and Welbl, Johannes and Gowal, Sven and Cherepanov, Alexey and Molloy, James and Mankowitz, Daniel J. and Sutherland Robson, Esme and Kohli, Pushmeet and de Freitas, Nando and Kavukcuoglu, Koray and Vinyals, Oriol},
  journal      = {Science},
  volume       = {378},
  number       = {6624},
  pages        = {1092--1097},
  year         = {2022},
  doi          = {10.1126/science.abq1158}
}

@misc{roziere2023codellama,
  title        = {Code Llama: Open Foundation Models for Code},
  author       = {Rozière, Baptiste and Gehring, Jonas and Gloeckle, Fabian and Sootla, Sten and Gat, Itai and Tan, Xiaoqing Ellen and Adi, Yossi and Liu, Jingyu and Remez, Tal and Rapin, Jérémy and others},
  year         = {2023},
  howpublished = {arXiv preprint arXiv:2308.12950},
  note         = {Accessed: 2025-07-22}
}

@inproceedings{pearce2022asleep,
  title        = {Asleep at the Keyboard? Assessing the Security of GitHub Copilot's Code Contributions},
  author       = {Pearce, Hammond and Ahmad, Baleegh and Tan, Benjamin and Dolan-Gavitt, Brendan and Karri, Ramesh},
  booktitle    = {Proceedings of the 2022 IEEE Symposium on Security and Privacy},
  pages        = {754--768},
  year         = {2022},
  note         = {arXiv:2108.09293}
}

@inproceedings{song2023empirical,
  title     = {An Empirical Study of Code Generation Errors Made by Large Language Models},
  author    = {Song, Chunqiu and Zhou, Mingyang and others},
  booktitle = {Workshop on Machine Learning for Programming (ML4P)},
  year      = {2023},
  url       = {https://mapsworkshop.github.io/assets/LLM_Code_Error_Analysis_MAPS2023_camera-ready.pdf}
}

@inproceedings{srikant2023generating,
  title        = {Generating Secure Code with Language Models},
  author       = {Srikant, Shashank and O'Brien, Benjamin and Tschiatschek, Sebastian and Bagdasaryan, Eugene and Rozen, Rishabh and Krishnamurthy, Vikash and others},
  booktitle    = {arXiv preprint arXiv:2306.23034},
  year         = {2023}
}

@inproceedings{tang2023codeagent,
  title        = {CodeAgent: Autonomous Programming with Conversational Software Agents},
  author       = {Tang, Mingxi and others},
  booktitle    = {arXiv preprint arXiv:2312.13010},
  year         = {2023}
}

@inproceedings{shrivastava2023ragcode,
  title        = {RAG-Code: Retrieval Augmented Generation for Code Synthesis},
  author       = {Shrivastava, Divyansh and Larochelle, Hugo and Tarlow, Daniel},
  booktitle    = {Proceedings of the 2023 Conference on Empirical Methods in Natural Language Processing},
  pages        = {2234--2245},
  year         = {2023}
}

@article{williams1992simple,
  title        = {Simple statistical gradient-following algorithms for connectionist reinforcement learning},
  author       = {Williams, Ronald J.},
  journal      = {Machine Learning},
  volume       = {8},
  number       = {3-4},
  pages        = {229--256},
  year         = {1992},
  publisher    = {Springer},
  doi          = {10.1007/BF00992696}
}

@inproceedings{demoura2008z3,
  title        = {Z3: An Efficient SMT Solver},
  author       = {de Moura, Leonardo and Bj{\o}rner, Nikolaj},
  booktitle    = {Tools and Algorithms for the Construction and Analysis of Systems},
  pages        = {337--340},
  year         = {2008},
  publisher    = {Springer},
  series       = {Lecture Notes in Computer Science},
  volume       = {4963},
  doi          = {10.1007/978-3-540-78800-3_24}
}

@misc{microsoft2020pylance,
  title        = {Pylance: Fast, feature-rich language support for Python in Visual Studio Code},
  author       = {{Microsoft Corporation}},
  year         = {2020},
  howpublished = {\\url{https://devblogs.microsoft.com/python/announcing-pylance-fast-feature-rich-language-support-for-python-in-visual-studio-code/}},
  note         = {Accessed: 2025-07-22}
}

@article{gupta1995maintenance,
  author       = {Gupta, Ashish and Mumick, Inderpal Singh},
  title        = {Maintenance of Materialized Views: Problems, Techniques, and Applications},
  journal      = {IEEE Data Engineering Bulletin},
  volume       = {18},
  number       = {2},
  pages        = {3--18},
  month        = {June},
  year         = {1995},
  note         = {Special Issue on Materialized Views and Data Warehousing}
}

@misc{bazel2015,
  title        = {Bazel: A Fast, Scalable, Multi-Language Build System},
  author       = {{Google Inc.}},
  year         = {2015},
  howpublished = {\\url{https://bazel.build/}},
  note         = {Open-sourced March 2015. Accessed: 2025-07-22}
}

@inproceedings{xu2022polycoder,
  title        = {A Systematic Evaluation of Large Language Models of Code},
  author       = {Xu, Frank F. and Alon, Uri and Neubig, Graham and Hellendoorn, Vincent J.},
  booktitle    = {Proceedings of the 6th ACM SIGPLAN International Symposium on Machine Programming (MAPS)},
  pages        = {1--10},
  year         = {2022},
  note         = {arXiv:2202.13169}
}

@inproceedings{husain2019codesearchnet,
  title        = {CodeSearchNet Challenge: Evaluating the State of Semantic Code Search},
  author       = {Husain, Hamel and Wu, Ho-Hsiang and Gazit, Tiferet and Allamanis, Miltiadis and Brockschmidt, Marc},
  booktitle    = {arXiv preprint arXiv:1909.09436},
  year         = {2019},
  note         = {Multi-language dataset covering Python, Java, JavaScript, PHP, Ruby, and Go}
}

@inproceedings{athiwaratkun2023multilingual,
  title        = {Multi-lingual Evaluation of Code Generation Models},
  author       = {Athiwaratkun, Ben and Gouda, Sanjay Krishna and Wang, Zijian and Li, Xiaopeng and Tian, Yuchen and Tan, Ming and Ahmad, Wasi Uddin and Wang, Shiqi and Sun, Qing and Shang, Mingyue and Gonugondla, Sujan Kumar and Ding, Hantian and Kumar, Varun and Fulton, Nathan and Farahani, Arash and Jain, Siddhartha and Giaquinto, Robert and Qian, Haifeng and Ramanathan, Murali Krishna and Nallapati, Ramesh and Ray, Baishakhi and Bhatia, Parminder and Sengupta, Sudipta and Roth, Dan and Xiang, Bing},
  booktitle    = {Proceedings of the 40th International Conference on Machine Learning (ICML)},
  year         = {2023},
  note         = {arXiv:2210.14868}
}

@article{cassano2023multipl,
  title        = {MultiPL-E: A Scalable and Polyglot Approach to Benchmarking Neural Code Generation},
  author       = {Cassano, Federico and Gouwar, John and Nguyen, Daniel and Nguyen, Sydney and Phipps-Costin, Luna and Pinckney, Donald and Yee, Ming-Ho and Zi, Yangtian and Anderson, Carolyn Jane and Feldman, Molly Q. and Guha, Arjun and Greenberg, Michael and Jangda, Abhinav},
  journal      = {IEEE Transactions on Software Engineering},
  volume       = {49},
  number       = {7},
  pages        = {3675--3691},
  year         = {2023},
  doi          = {10.1109/TSE.2023.3267446},
  note         = {arXiv:2208.08227}
}

\end{document}